\documentclass[12pt,preprint]{aastex}

\usepackage{amsmath}

\newcommand{\dd}{\ensuremath{\mathrm{d}}}

\shorttitle{Turbulence in a 3D deflagration model for type Ia SNe: 
  II. Intermittency and DDT probability}
\shortauthors{Schmidt et al.}

\begin{document}


\title{Turbulence in a three-dimensional deflagration model for type Ia supernovae: 
  II. Intermittency and the deflagration-to-detonation transition probability}


\author{W. Schmidt}
\affil{Institut f\"ur Astrophysik, Universit\"at G\"ottingen, Friedrich-Hund-Platz 1,
D-37077 G\"ottingen, Germany
\and Lehrstuhl f\"ur Astronomie und Astrophysik, Universit\"at W\"urzburg, Am Hubland, D-97074 W\"urzburg, Germany}
\email{schmidt@astro.physik.uni-goettingen.de}

\author{F. Ciaraldi-Schoolmann}
\affil{Max-Planck-Institut f\"ur Astrophysik, Karl-Schwarzschild-Str. 1, D-85741 Garching, Germany\and Lehrstuhl f\"ur Astronomie und Astrophysik, Universit\"at W\"urzburg, Am Hubland, D-97074 W\"urzburg, Germany}

\author{J. C. Niemeyer}
\affil{Institut f\"ur Astrophysik, Universit\"at G\"ottingen, Friedrich-Hund-Platz 1,
D-37077 G\"ottingen, Germany} 

\author{F. K. R\"opke and W. Hillebrandt}
\affil{Max-Planck-Institut f\"ur Astrophysik, Karl-Schwarzschild-Str. 1, D-85741 Garching, Germany}



\begin{abstract}
The delayed detonation model describes the observational properties of
the majority of type Ia supernovae very well. Using numerical data
from a three-dimensional deflagration model for type Ia supernovae,
the intermittency of the turbulent velocity field and its implications
on the probability of a deflagration-to-detonation (DDT) transition
are investigated. From structure functions of 
the turbulent velocity fluctuations, we determine intermittency
parameters based on the log-normal and the log-Poisson models. 
The bulk of turbulence in the ash regions appears to be less intermittent 
than predicted by the standard log-normal model and the She-L\'{e}v\^{e}que model.
On the other hand, the analysis of the turbulent velocity fluctuations in the vicinity 
of the flame front by R\"opke suggests a much higher probability of large 
velocity fluctuations on the grid scale in comparison
to the log-normal intermittency model. Following Pan et al., we computed
probability density functions for a DDT for the different distributions.
The determination of the total number of regions at
the flame surface, in which DDTs can be triggered, enables us to estimate
the total number of events. Assuming that a DDT can occur in the stirred flame regime, 
as proposed by Woosley et al., the log-normal model would imply a 
delayed detonation between $0.7$ and $0.8$ seconds after the beginning of the deflagration
phase for the multi-spot ignition scenario used in the simulation. However, the
probability drops to virtually zero if a DDT is further constrained by
the requirement that the turbulent velocity fluctuations reach about $500\,\mathrm{km\,s^{-1}}$. 
Under this condition, delayed detonations are only
possible if the distribution of the velocity fluctuations is not log-normal.
From our calculations follows that the distribution obtained by R\"opke allow 
for multiple DDTs around $0.8$ seconds after ignition at a transition density close to 
$1\times 10^{7}\,\mathrm{g\,cm^{-3}}$.
\end{abstract}





\section{Introduction}
\label{Intro}

Although deflagration models of Type Ia supernovae
\citep[e.g.][]{Rein02, Gam03, Roep05} reproduce bulk properties of
fainter Type Ia supernovae, delayed detonations \citep[see][for a
review of explosion scenarios]{HilleNie00} seem to be
the the most promising way of modeling the majority of the observed
events \citep{RoepNie07, Mazzali07}.

However, a theoretical explanation for the
deflagration-to-detonation (DDT) transition that gives rise to this
type of thermonuclear explosion has been missing so far.   
Originally
postulated by \citet{Khokhlov3} and \citet{WoosWeav94},
the possibility of a DDT in type Ia
supernovae was questioned by \citet{Niemeyer}, because the
preconditioning required to trigger a self-sustaining detonation front
would be extremely unlikely to occur in the deflagration
phase. This conclusion was based on ensemble-average scaling
arguments. 
\citet{Pan}, on the other hand, pointed out that highly
intermittent turbulent fluctuations act in favour of a
DDT, because intermittency allows for velocities much
higher than the ensemble average. 

As proposed by \citet{Niemeyer2}, \citet{Niemeyer3} and \citet{Khokhlov4}, 
a necessary condition for triggering a DDT is that the distributed burning regime has to be entered. 
Based on log-normal \citep{Kolmogorov} and log-Poisson \citep{She} intermittency models
with typical parameters expected for turbulence in thermonuclear
supernovae, Pan et al. calculated probabilities for this DDT condition.
Their calculations implied transition densities in the range $3.8\times 10^{7}\ldots 2.7\times
10^{8}\,\mathrm{g\,cm^{-3}}$. \citet{LisHille00}, \citet{Woosley07},
and \citet{WoosKer09}, on the other hand, provided numerical evidence that there are stronger constraints on a DDT than just entering the regime of distributed burning. As a consequence, lower transition densities would be favored. Analyzing the small-scale turbulent velocity distribution
predicted by the subgrid-scale model in several thermonuclear supernova simulations, \citet{Roep07} found that intermittent velocity fluctuations up to $1000\,\mathrm{km\,s^{-1}}$ on a
length scale of $10\,\mathrm{km}$ are possible. Fluctuations of this magnitude could trigger a detonation in accordance with \citet{LisHille00}.

The ambiguity of conditions for DDTs in type Ia supernovae makes predictions difficult. Here, we follow a pragmatic approach and investigate different DDT criteria, assuming simple approximations to the microphysical conditions. Following the approach of Pan et al., we calculate the probability of
a DDT using data sets from a highly resolved three-dimensional deflagration model
\citep{Roepke}. \citet{CirSchm08}, in the following referred to as
paper I, determined characteristic scales and scaling exponents from the structure functions
of the turbulent velocity fluctuations up to order six. We fitted
log-normal and log-Poisson intermittency models to these scaling exponents.
In addition, we determined probability density functions (pdfs) of the mass density constrained
to the flame surface in order to calculate the probability of a DDT at different stages of the
deflagration. There are no further adjustable parameters apart from the initial conditions of the
deflagration model. Assuming a certain size for DDT regions, the total number of DDTs can be obtained from the cumulative probability and the number of regions in the volume occupied by the flame.\footnote{
Down to the Gibson scale, the flame front can be considered as a fractal surface, which is numcerically represented by
a zero level set. Thus, the volume occupied by the flame is defined by all grid cells within a certain
distance to the zero level set.}

Basically, \citet{Pan} assumed that a DDT is triggered once the Karlovitz
number $\mathrm{Ka}$ (the square root of the ratio of the laminar
flame width to the Gibson scale as defined in Section~\ref{sc:ddt_prob}) exceeds a certain value
\citep{Niemeyer3,WoosKer09}, i.~e., burning has advanced sufficiently far into the
regime of distributed burning.  Because the critical size of a preconditioned region
is extremely sensitive to the mass density, it is difficult to estimate the total number of DDTs.
Nevertheless, our results suggest that a delayed detonation would result at an early
stage of the explosion if the conditions for a DDT were constrained by the Karlovitz
number only. A further implication would be a transition density well above 
$1\times 10^{7}\,\mathrm{g\,cm^{-3}}$, which was found in the study by \citet{Pan}. 

Whereas the criterion for the onset of distributed burning can be expressed in terms of laminar flame
properties, \citet{WoosKer09} argued that the conditions for a DDT cannot 
be parameterized on the basis of the laminar flame speed and thickness. 
Using numerical data for the nuclear time scale in the so-called stirred flame regime \citep{Kerst01}
and following the same procedure as outlined above, we calculated the DDT probability.
Although there are substantial uncertainties in this approach, a DDT at later time and lower density
compared to the estimates following from laminar flame properties
is implied. Most important, we find that a DDT can be excluded if the
magnitude of the turbulent velocity fluctuations is required to exceed about
$500\,\mathrm{km\,s^{-1}}$, which is suggested by microphysical studies
\citep{Woosley07}. Delayed detonations are possible, however, if the distribution of
the velocity fluctations significantly deviates from a log-normal shape. 
According to \citet{Roep07}, this appears to be the case for regions near the flame front. 

The paper is organized as follows. In the next Section, we briefly
review the log-normal and log-Poisson intermittency models. Moreover, the
results on the intermittency of turbulence in the simulation are discussed. The
procedure of calculating the DDT probability is described in Section~\ref{sc:ddt_prob}.
In Section~\ref{sc:number_ddt}, the estimation of the total number of DDTs is explained.
We present our numerical results in this Section~\ref{sc:results}, followed by a discussion in the last Section.


\section{Intermittency}

According to the Kolmogorov theory of statistically stationary,
homogeneous, and isotropic turbulence, the structure functions
$S_{p}(\ell):=\langle\delta v^{p}(\ell)\rangle$, i.~e., the ensemble
average of the velocity fluctuations to the power $p$ over a length
scale $\ell$, follow power laws in the inertial subrange. The scaling
exponents are given by $\zeta_{p}=p/3$. In several experiments,
however, departures from the $p/3$ scaling were found for higher-order
structure functions. This phenomenon is attributed to the
intermittency of turbulence \citep[][see]{Frisch}. There are various
theoretical intermittency models that predict anomalous scaling
exponents. One example is the Kolmogorov-Oboukhov model
\citep{Kolmogorov, Oboukhov}, which assumes a log-normal probability
distribution function for the turbulent energy dissipation $\epsilon_{\ell}$
on a length scale $\ell$:
\begin{equation}
  \mathrm{pdf}\left(\ln(\epsilon_{\ell}/\epsilon)\right) =
  \frac{1}{\sqrt{2\pi\sigma_{\ell}^{2}}}
  \exp\left(-\frac{\left[\ln(\epsilon_{\ell}/\epsilon)+\sigma_{\ell}^{2}/2\right]^{2}}{2\sigma_{\ell}^{2}}\right),
  \label{eq:lognorm}
\end{equation}
where $\epsilon$ is the mean rate of energy dissipation and
$\sigma_{\ell}^2=\mu\ln(L/\ell)$ for the integral scale $L$.
The scaling exponents resulting from this model are given by
\begin{equation}
 \zeta_{p}^{(\mathrm{ln})}=\frac{p}{3}-\frac{1}{18}\mu p\left(p-3\right),
\label{Eq3}
\end{equation}
where the parameter $\mu$ is defined by $\mu = 2-\zeta_{6}$. Experiments
and simulations indicate $\zeta_{6}\approx 1.8$ and therefore
$\mu\approx 0.2$. A defficiency of the log-normal model is that it fails
to describe the observed higher-order scalings for $p\ge 10$ (see
\citet{Pan} for a detailed discussion of the limitations of this
model). In this respect, the intermittency model by \cite{She} has
been particularly successful:
\begin{equation}
\zeta_{p}^{(\mathrm{SL})}=\frac{p}{9}+2\left[1-\left(\frac{2}{3}\right)^{p/3}\right].
\label{Eq4}
\end{equation}
\cite{Dubr} and \cite{She2} showed that this model can be accommodated
within a two-parameter family of intermittency models, which are based
on log-Poisson statistics:
\begin{equation}
\zeta_{p}^{(\mathrm{lP})} =(1-\Delta)\frac{p}{3}+\frac{\Delta}{1-\beta}\left(1 - \beta^{p/3}\right)
\label{Eq4b} 
\end{equation}
The random cascade factor $\beta\in[0,1]$ specifies the degree of
intermittency. Non-intermittent Kolmogorov scaling is obtained in the
limit $\beta\rightarrow 1$. The parameter $\Delta$ specifies the
scaling properties of the most intermittent dissipative structures and
is related to the codimension $C$ of these structures:
$C=\Delta/(1-\beta)$. Setting $\Delta=2/3$ and $\beta=2/3$, the
She-L\'{e}v\^{e}que model~(\ref{Eq4}) follows from
equation~(\ref{Eq4b}).  For these parameters, the codimension is
$C=2$. The interpretation is that vortex filaments are the most
dissipative structures in subsonic turbulence \citep{She}. For
supersonic turbulence, $1\le C<2$ because shock fronts also
dissipate energy \citep{Boldyrev,KritPad07,SchmFeder09}.

\begin{figure}
        \plottwo{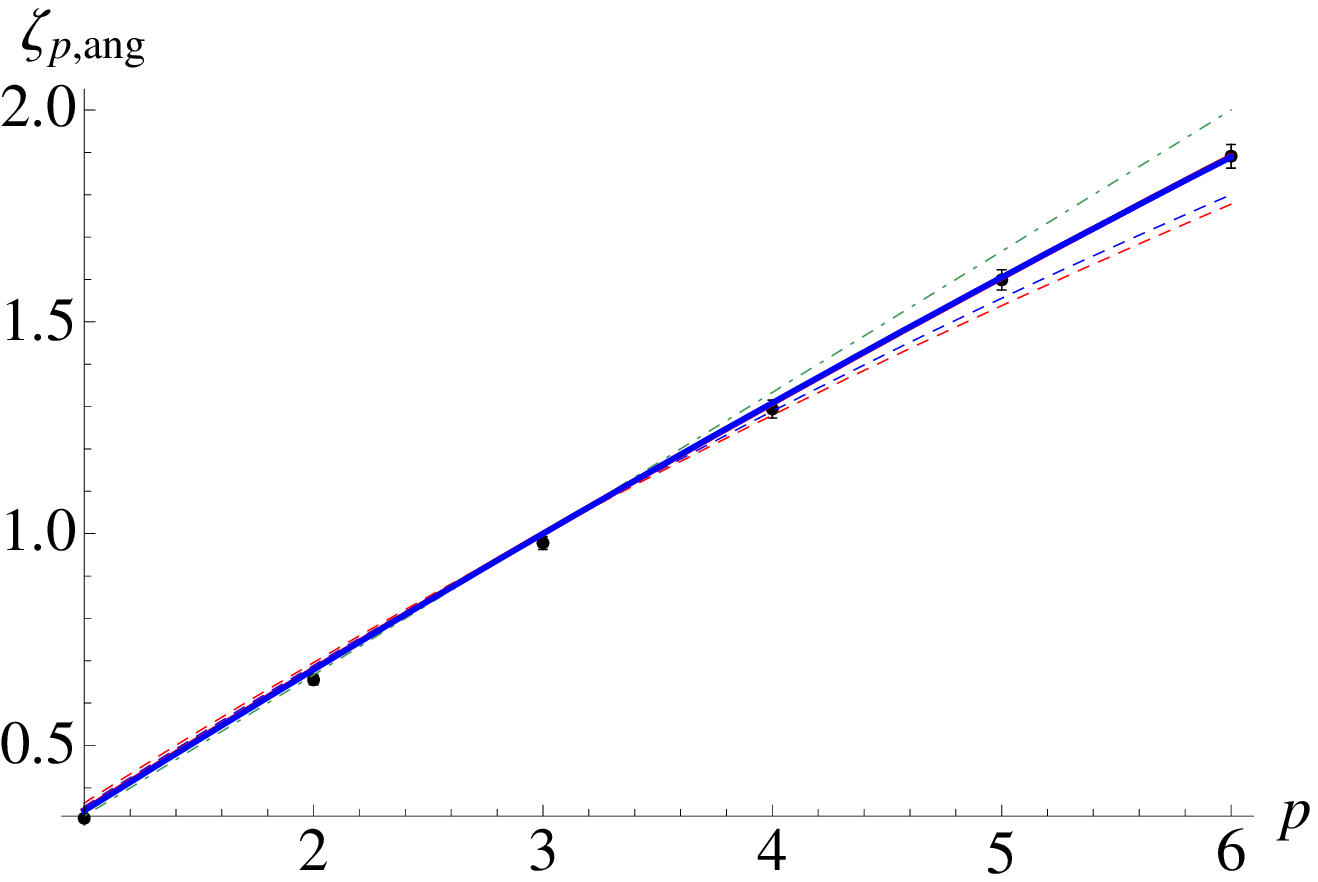}{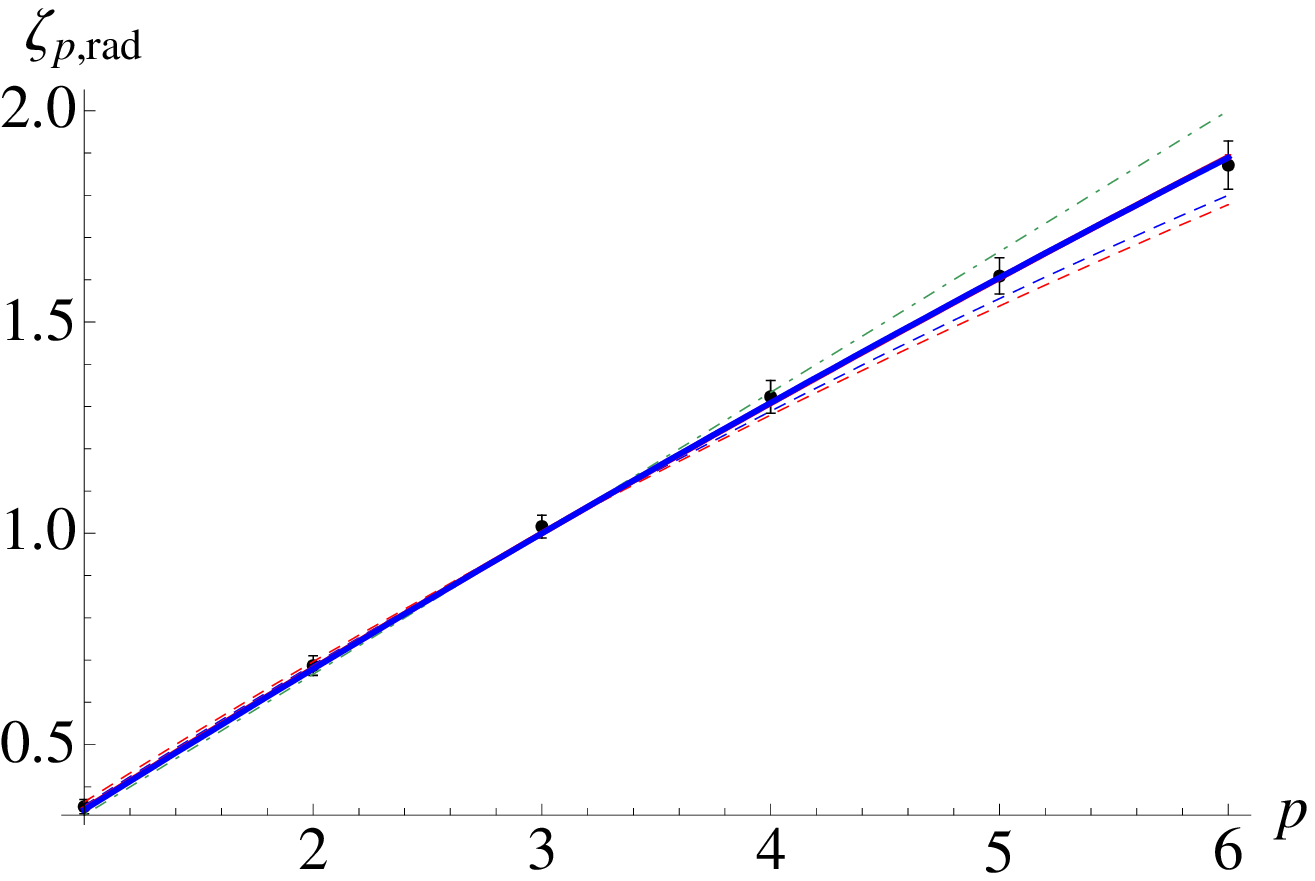}
        \caption{Scaling exponents of the angular (left) and radial (right) structure functions of order $p\le 6$ at time $t = 0.7\,\mathrm{s}$. For comparison, the predictions of the
        Kolmogorov theory (dot-dashed line), the log-normal model (blue dashed line in the online version) with $\mu=0.2$ and the She-Leveque model (red dashed line in the online version) are plotted. The solid lines show the log-normal and log-Poisson fit functions, which are nearly coinciding. For the log-Poisson model, we set the codimension $C=2$. The corresponding fit parameters are listed in Table~\ref{parameters}.}
         \label{fig:INT}
\end{figure}

Equation~(\ref{Eq3}) as well as~(\ref{Eq4}) imply $\zeta_3=1$ and,
hence, $S_{3}(\ell)\sim \epsilon\ell$. In paper I, we determined the
scaling exponents form power-law fits to the structure functions for
$p\le 6$. It was found that there is a transition length 
$\ell_{\mathrm{K/RT}}\sim 10\,\mathrm{km}$, 
where the radial second-order structure
function changes from Kolmogorov scaling ($S_{2}(\ell)\propto \ell^{2/3}$) to Rayleigh-Taylor scaling
($S_{2}(\ell)\propto \ell$). The values of $\ell_{\mathrm{K/RT}}$ for $t\ge 0.5$ seconds
are listed in Table~\ref{parameters}.
While $\zeta_{3}$ is approximately unity for the
angular structure functions and also for the radial structure
functions in the subrange $\ell<\ell_{\mathrm{K/RT}}$, the radial third-order
exponent for $\ell>\ell_{\mathrm{K/RT}}$ is about $1.5$. On the other hand,
it was shown in paper I that the relative scaling exponents $Z_{p}=\zeta_{p}/\zeta_{3}$
of the radial structure functions are nearly equal for the two subranges. However, since a
consistent treatment of an exponent $\zeta_{3}$ different from unity
is highly non-trivial in the framework of the intermittency models
mentioned above \citep{Dubr,SchmFeder08}, we do not consider the
radial scaling exponents for length scales greater than
$\ell_{\mathrm{K/RT}}$. The values of $\mu$ resulting from fits of the
log-normal model to the exponents $\zeta_{p}$ for
$t=0.5,\ldots,0.9$ seconds are summarized in Table~\ref{parameters}. 
Table~\ref{zeta_6} demonstrates that these values are consistent
with $\mu=2-\zeta_{6}$ within the error bars of the
scaling exponents. For $t=0.7$ seconds,
the scaling exponents of the radial and angular structure
functions are plotted together with the fit
functions in Figure~\ref{fig:INT}. For comparison, also the log-normal
model with $\mu=0.2$ and the She-Leveque model~(\ref{Eq4}) are
plotted. We find $\mu\approx 0.1$.
Clearly, the numerical data from the supernova simulation
indicate a lower degree of intermittency than these two models.
Moreover, there is a remarkable agreement between the radial scaling exponents
for the range of length scales $\ell<\ell_{\mathrm{K/RT}}$ and
the scaling exponents of the angular structure functions over the whole 
range of length scales (also see paper I).

Fitting the general log-Poisson model~(\ref{Eq4}) to the scaling exponents, yields ambiguous results. \citet{Pan} propose to set $\Delta=2/3$. This appears to be a sensible choice on grounds of the hypothesis of a universal
dissipation time scale, which determines $\Delta$. However, the best
fits constrained by $\Delta=2/3$ imply a codimension greater than $2$. Since
it is hard to come up with a reasonable physical interpretation of this result,
we tested the hypothesis $C=2$, i.~e., we consider the codimension of the most intense dissipative structure to be fixed. The resulting fit functions are almost indistinguishable from
the best log-normal fits (see Figure~\ref{fig:INT}). The log-Poisson fit parameters are also listed in
Table~\ref{parameters}.  As one can see, $\beta>2/3$, indicating a
lower degree of intermittency in comparison to the She-Leveque model,
and $\Delta\approx 0.5$. In this respect, the trend indicated by
the log-normal model fits is confirmed. The anomalous intermittency
parameters point at deviations from fully developed turbulence. A
possible reason is that turbulence in supernova explosions does not
reach a statistically stationary state. Since there are indications for $\Delta>2/3$
in the case of highly compressible turbulence \citep{SchmFeder08,SchmFeder09}, the lower
value of $\Delta$ in the present case cannot be attributed to compressibility. 
Since the log-Poisson model is sensitive even to small errors in the
higher-order scalings, the deviations from the She-Leveque model might
be spurious.  In this regard, however, one has to keep in mind that
insufficiently converged statistics usually causes an \emph{over}estimate of
intermittency rather than an underestimate \citep[see][]{Frisch}. 

Assuming that the low value of $\mu$ is a peculiar property of
non-stationary RT-driven turbulence in thermonuclear supernovae,
one might raise the question if the small-scale isotropy discussed in paper I
is genuine or, possibly, an artifact of the employed SGS model.
Since our SGS model features a localized closure for the turbulence
energy flux \citep{SchmNie06a}, isotropy on numerically resolved scales
is not required. Only the unresolved turbulent velocity fluctuations
are assumed to be isotropic. Fig. 2 in paper I demonstrates that 
the resolved velocity field is isotropic on length scales smaller
than $\ell_{\mathrm{K/RT}}$. At late time, when
$\ell_{\mathrm{K/RT}}$ becomes less than the size of the
grid cells, the radial and angular structure functions 
deviate even on the smallest numerically resolved scales
and, thus, we have anisotropy on these scales.
Consequently, isotropy does not appear to be forced
by the SGS model if there is a pronounced
anisotropy on length scales near the numerical cutoff scale.
Although we cannot exclude the possibility that the SGS
tends to make the nearby resolved scales more isotropic than in reality,
because the interactions between resolved and unresolved scales
are not strictly local, it is plausible that the observed isotropy on length scales
$\ell\lesssim\ell_{\mathrm{K/RT}}$ is basically genuine.

Analyzing the statistics of the SGS velocity fluctuations in the
vicinity of the flame front, \citet{Roep07} found the high-velocity tail of the pdf is very well 
fitted by the expression $\mathrm{pdf}(\delta v(\ell))=\exp\left[{a_{0}\delta v^{a_{1}}(\ell)+a_{2}}\right]$ for $\delta v(\ell)> 10^{7}\,\mathrm{cm\,s^{-1}}$, where $\ell=\Delta$. The local velocity
fluctuation at the numerical cutoff is given by the specific SGS turbulence energy:
$\delta v(\Delta)=q_{\mathrm{sgs}}=\sqrt{2k_{\mathrm{sgs}}}$ \citep{SchmNie06a}.
The coefficients $a_{0}$, $a_{1}$, and $a_{2}$ are fitted for each instant separately. The variation of
the coefficients is due to the evolution of the turbulent flow. Changes might also result from the
temporal shift of the cutoff scale. For the phase of the explosion we are interested in, $\Delta$ is about $10\,\mathrm{km}$, which in turn is close to $\ell_{\mathrm{K/RT}}$. Further studies indicated that the above form of the pdf also holds for resolved velocity fluctuations on length scales $\ell\sim\ell_{\mathrm{K/RT}}$.
Thus, we assume that the fit coefficients do not change significantly
for nearby scales at a fixed time. The corresponding pdf for $\epsilon_{\ell}=\delta v^{3}(\ell)/\ell$ is given by
\begin{equation}
  \mathrm{pdf}(\epsilon_{\ell}) =
  \frac{\ell^{1/3}}{3\epsilon_{\ell}^{2/3}}
  \exp\left[a_{0}(\ell\epsilon_{\ell})^{a_{1}/3}+a_{2}\right] \qquad
  \mbox{for}\ \ell\sim10\,\mathrm{km}.
  \label{eq:roepke}
\end{equation}
Fig.~1 in  \citet{Roep07} shows that the high-velocity tail is much flatter than the
prediction of the log-normal intermittency model. Consequently, there appears
to be a much higher degree of intermittency. It is important
to note that this applies to velocity fluctuations at the flame surface. In this regard,
the behavior of turbulence close to the flame front appears to be markedly different from the
properties of the bulk of turbulence in ash regions, for which the log-normal fits to the
scaling exponents imply a relatively \emph{low} degree of intermittency. We will see
in Section~\ref{sc:results} that this difference is crucial for the occurrence of delayed detonations if
tight constraints on DDTs are assumed.

\section{Deflagration-to-detonation transition probability}
\label{sc:ddt_prob}

Turbulent deflagration in the flamelet regime is characterized by a
flame thickness $\ell_{\mathrm{fl}}<\ell_{\mathrm{G}}$
\citep{Niemeyer3}, where the Gibson scale $\ell_{\mathrm{G}}$ is the
length scale for which the turbulent velocity fluctuations are
comparable to the laminar flame speed: $\delta
v(\ell_{\mathrm{G}})\sim s_{\mathrm{lam}}$. According to the refined
similarity hypothesis of \citet{Kolmogorov}, we have $\delta
v(\ell)\sim \epsilon^{1/3}\ell^{1/3}$. Assuming isotropic turbulence,
the mean rate of dissipation $\epsilon=V^{3}/L$, where $L$ and $V$ are
the integral length and the associated velocity,
respectively. However, it was shown in paper I that turbulence in
thermonuclear supernovae is anisotropic on length scales
considerably smaller than $L$. We will therefore give
a more precise definition below. In the ensemble average, it follows
from $\ell_{\mathrm{G}}=(s_{\mathrm{lam}}^{3}/V)^{3}L$ that
$\epsilon \sim \mathrm{Ka}^2\epsilon_{\mathrm{fl}}$, where
$\epsilon_{\mathrm{fl}}:= s_{\mathrm{lam}}^{3}/\ell_{\mathrm{fl}}$ and
the Karlovitz number
$\mathrm{Ka}=(\ell_{\mathrm{fl}}/\ell_{\mathrm{G}})^{1/2}$
\citep{Peters00}. Once $\ell_{\mathrm{G}}$ becomes smaller than
$\ell_{\mathrm{fl}}$, i.~e., $\mathrm{Ka}>1$, distributed burning will
commence \citep{Niemeyer2,Khokhlov4}. It should be noted that
the meaning of $\mathrm{Ka}$ in terms of the Gibson length becomes
elusive in the limit $\mathrm{Ka}\gg1$, because the flame thickness and
the laminar burning speed are not well defined in this regime.

\subsection{Estimation of the DDT probability based on laminar flame properties}
\label{sc:DDTlam}

In log-normal as well as log-Poisson intermittency models, the rate of
dissipation averaged over a region of size $\ell$ is defined by a
random variable $\epsilon_{\ell}$ with a certain probability density
function $\mathrm{pdf}(\epsilon_{\ell})$. As pointed out by
\citet{Pan}, the probability of distributed burning with 
a certain Karlovitz number in a region of size $\ell$ is then given by\footnote{They use $K=\mathrm{Ka}^{2/3}$ as a fudge parameter, without referring to the Karlovitz number.}
\begin{equation}
P(\epsilon_{\ell} > \mathrm{Ka}^2\epsilon_{\mathrm{fl}})=\int_{\mathrm{Ka}^2\epsilon_{\mathrm{fl}}}^{\infty}\,\mathrm{pdf}(\epsilon_{\ell}')\,\dd\epsilon_{\ell}'.
\end{equation}
As the Karlovitz number, for which a DDT can occur, increases, the probability given by the
above equation declines. Apart from $\mathrm{Ka}$, the threshold for $\epsilon_{\ell}$ is given by
laminar flame properties, because $\epsilon_{\mathrm{fl}}$ is defined by the speed and width of laminar flames.

Assuming that a DDT is caused by the Zel'dovich mechanism, there must be a
sufficient number of preconditioned regions that reach the critical
size $\ell_{\mathrm{c}}$. This length scale depends on the mass
density and the composition of the fuel \citep{Niemeyer2}. While
\citet{Pan} assumed a fixed density and a spherical flame, we evaluate
the simulation data described in paper I. Specifically, we compute
the effective probability that a single DDT occurs anywhere near the flame front
from the convolution of the conditional probability $P(\epsilon_{\ell} > \mathrm{Ka}^2\epsilon_{\mathrm{fl}})$ with the normalized
probability density function of the mass density $\rho$ constrained to the flame front:
\begin{equation}
\label{eq:prob_ddt_ka}
P_{\mathrm{DDT}}(\mathrm{Ka})=\int_{0}^{\infty}P(\epsilon_{\ell_{\mathrm{c}}} > \mathrm{Ka}^2\epsilon_{\mathrm{fl}})\,\mathrm{pdf}(\rho|G=0)\,\dd\rho.
\end{equation}
The flame front is numerically defined by the zero level set, $G=0$,
where $G$ denotes the the level set function. Basically,
$G(t,\mathbf{r})$ is a distance function that is positive inside the
flame (in ash regions) and negative outside. For the
numerical calculation of $\mathrm{pdf}(\rho|G=0)$, the grid cells
in which $G(t,\mathbf{r})$ switches sign are identified. 
The distribution of the mass density over the flame surface
is mostly due to the density stratification in the exploding star.
Since turbulence in SNe Ia is weakly compressible, we ignore
correlations between local fluctuations of the density and the
velocity.

For the calculation of $P(\epsilon_{\ell_{\mathrm{c}}} >
\mathrm{Ka}^2\epsilon_{\mathrm{fl}})$, the probability density
function of $\epsilon_{\mathrm{\ell}}$ has to be modelled. 
From the log-normal probability density function~(\ref{eq:lognorm}), 
the following expression for $P(\epsilon_{\ell_{\mathrm{c}}} >
\mathrm{Ka}^2\epsilon_{\mathrm{fl}})$ is obtained \citep{Pan}:
\begin{equation}
  P\left(\epsilon_{\ell_{\mathrm{c}}} > \mathrm{Ka}^2\epsilon_{\mathrm{fl}}\right) = \frac{1}{2}\mathrm{erfc}\left(\frac{\ln\left(\mathrm{Ka}^2\epsilon_{\mathrm{fl}}/\epsilon\right)}{\sqrt{2}\sigma_{\ell_{\mathrm{c}}}}+\frac{\sigma_{\ell_{\mathrm{c}}}}{2\sqrt{2}}\right)
  \label{eq:prob_ddt_ka_tot}
\end{equation}
For the log-Poisson model, on the other hand, it is not possible to derive an analytic expression for
$P\left(\epsilon_{\mathrm{c}} > \mathrm{Ka}^2\epsilon_{\mathrm{fl}}\right)$, because the Poisson distributions contributing to $\mathrm{pdf}(\epsilon_{\ell})$ are not explicitly known. For this reason, we will not consider the log-Poisson intermittency model in the following. 

The restrictions of the log-normal model were discussed at length by \citet{Pan}. Data from high-resolution simulations of turbulence suggest that the log-normal model yields a good approximation to the distribution of $\epsilon_{\ell}$
within the $5\sigma_{\ell}$ wings. However, it is not clear whether this range also
applies to turbulence in thermonuclear supernova. A conservative estimate can be made as follows: 
Given a random variable $X$, the largest $X$-values, for which 
the probability density function $\mathrm{pdf}(X)$ is constrained by the moments of order $\le p$,
is indicated by the peak of $X^{p}\mathrm{pdf}(X)$. Since $\epsilon_{\ell}\sim \delta v^{3}(\ell)/\ell$, we have 
$\delta v^{p}(\ell)\mathrm{pdf}(\delta v(\ell))\sim
\delta v^{p-1}(\ell)\mathrm{pdf}\left(\ln[\delta v^{3}(\ell)/\ell\epsilon]\right)$,
where the pdf on the right-hand side is defined by equation~(\ref{eq:lognorm}).
We computed the scaling exponents of the structure functions up to sixth order.
For $\ell=10\,\mathrm{km}$ and the numerical parameters listed in Table~\ref{parameters}, 
$\delta v^{6}(\ell)\mathrm{pdf}(\delta v(\ell))$ peaks for 
velocity fluctuations around $100\,\mathrm{km\,s^{-1}}$, which corresponds to a range within $2\sigma_{\ell}$ from the maximum of the log-normal $\epsilon_{\ell}$-pdf. Consequently, the model extrapolates the central part of the pdf, which can be inferred from the known structure functions, to the far tail. It is important to note that
the range, for which the log-normal pdf is not well constrained by the scaling exponents, is about the range
of the asymptotic tail~(\ref{eq:roepke}) resulting from the analysis of small-scale velocity fluctuations by \citet{Roep07}. In order to compare to the results of \citet{Pan}, however, we will also consider a log-normal shape of the pdf up to $5\sigma_{\ell}$.

For the determination of the mean dissipation rate $\epsilon$, it is important to bear in mind
that intermittency models such as the log-normal model apply to
statistically isotropic turbulence. It was shown in paper I, that
Kolmogorov scaling applies on length scales
$\ell < \ell_{\mathrm{K/RT}}$. For this reason, it is not correct
to set $\epsilon=V^{3}/L$, where $V$ and $L$ are the scales of energy
injection, in the case of RT-driven turbulence. In fact, the rate of
dissipation is fixed by the transition length $\ell_{\mathrm{K/RT}}$
and the velocity scale $\delta v(\ell_{\mathrm{K/RT}})$. 
Since the third-order structure function $S_{3}(\ell)\simeq
\epsilon\ell$, we define
$\epsilon_{\mathrm{rad}}:=S_{3,\mathrm{rad}}(\ell_{\mathrm{K/RT}})/\ell_{\mathrm{K/RT}}$. However, one can also set $\epsilon_{\mathrm{ang}}:=S_{3,\mathrm{ang}}(L)/L$, because
the scaling of the angular structure function is unique and consistent with the Kolmogorov theory for the whole range of length scales. Table~\ref{parameters} gives an overview of the numerical values of 
$\epsilon_{\mathrm{rad}}$, $\epsilon_{\mathrm{ang}}$,
and $L$ as functions of time.
While both definitions yield similar results for
the mean rate of dissipation, it is obvious from Figure 3 in paper I
that $\epsilon \ll V^{3}/L \sim S_{3,\mathrm{rad}}(L)/L$ if $V$ is
taken to be the characteristic velocity of the Rayleigh-Taylor-driven turbulent flow at the
length scale $L$. On the other hand, it is essential to set the integral length scale
for the distribution of $\epsilon_{\ell}$ to $L$ rather than $\ell_{\mathrm{K/RT}}$,
because intermittent velocity fluctuations occur on all length scales $\ell<L$.
Although turbulence is anisotropic for $\ell_{\mathrm{K/RT}}\lesssim\ell\lesssim L$, we assume that the log-normal model can be applied to the whole range of length scales
on the basis of the dissipation rate $\epsilon_{\mathrm{ang}}\simeq 
\epsilon_{\mathrm{rad}}$, because the angular structure functions continue to follow the scaling laws of isotropic turbulence for $\ell > \ell_{\mathrm{K/RT}}$. In Section~\ref{sc:results}, we will
carry out numerical calculations for both the radial and the angular parameter sets.

\subsection{Estimation of the DDT probability in the stirred flame regime}

So far, we have assumed that a DDT is only constrained by a value of $\mathrm{Ka}$
greater than unity. However, \citet{WoosKer09} argued that an additional
requirement for a DDT is that $\mathrm{Da} > 1$, where the Damk\"{o}hler number 
$\mathrm{Da}=T/\tau_{\mathrm{nuc}}$ is the ratio of a dynamical time scale $T$ to the nuclear burning time scale $\tau_{\mathrm{nuc}}$  \citep{Kerst01}. This regime is called
the stirred flame (SF) regime. In the LEM study by \citet{WoosKer09}, it was found that
the size of a region, in which a detonation can be triggered, is about $10\,\mathrm{km}$.
Since the values of $\ell_{\mathrm{K/RT}}$ listed Table~\ref{parameters}
are comparable to this size, we assume that $\mathrm{Da}\sim 1$ corresponds
to a burning zone of size $\sim\ell_{\mathrm{K/RT}}$. Then $T=\ell_{\mathrm{K/RT}}/\delta v(\ell_{\mathrm{K/RT}})$ is the turn-over time of the largest eddies in the nearly isotropic
regime. Since Kolmogorov scaling applies on length scales $\ell \lesssim
\ell_{\mathrm{K/RT}}$, it follows that $T=\ell_{\mathrm{K/RT}}^{2/3}/\epsilon$. For $\mathrm{Da} > 1$, there are broadened flame structures of size smaller than $\ell_{\mathrm{K/RT}}$.

In the framework of intermittency theory, the Damk\"{o}hler number in a region of size $\ell_{\mathrm{K/RT}}$ is given by 
\begin{equation}
	\mathrm{Da}=
	\frac{\ell_{\mathrm{K/RT}}^{2/3}}{\epsilon_{\ell_{\mathrm{K/RT}}}^{1/3}\tau_{\mathrm{nuc}}},
\end{equation}
where $\epsilon_{\ell_{\mathrm{K/RT}}}$ is interpreted as the random dissipation rate
on the length scale $\ell_{\mathrm{K/RT}}$. Hence, the requirement $\mathrm{Da} > 1$ for a DDT in the SF regime corresponds to an \emph{upper} bound on $\epsilon_{\mathrm{K/RT}}$:
\begin{equation}
	\label{eq:eps_max}
	\epsilon_{\ell_{\mathrm{K/RT}}} < \epsilon_{\mathrm{WSR}}:=\frac{\ell_{\mathrm{K/RT}}^{2}}{\tau_{\mathrm{nuc}}^{3}}.
\end{equation}
If $\mathrm{Da}$ becomes less than unity ($\epsilon_{\ell_{\mathrm{K/RT}}} \ge \epsilon_{\mathrm{WSR}}$), the well stirred reactor (WSR) regime is entered. In this regime, the density
becomes so low that a detonation is very difficult and, eventually, the flames will be quenched.

On the other hand, if $\mathrm{Da}$ is yet too high, the broadened flames produced at the onset of distributed burning will be too sparse to coalesce into a mixed flame structure extending over $\ell_{\mathrm{K/RT}}$ \citep{WoosKer09}. For this reason, there is a \emph{lower} bound on $\epsilon_{\mathrm{K/RT}}$ corresponding to a critical Damk\"{o}hler number $\mathrm{Da}_{\,\mathrm{crit}}$:
\begin{equation}
	\label{eq:eps_min}
	\epsilon_{\ell_{\mathrm{K/RT}}} > \epsilon_{\mathrm{crit}}:=
	\frac{\ell_{\mathrm{K/RT}}^{2}}{\mathrm{Da}_{\,\mathrm{crit}}^{3}\tau_{\mathrm{nuc}}^{3}}.
\end{equation}
If a DDT was only constrained by the range of $\mathrm{Da}$, in principle, very small velocity
fluctuations could trigger a delayed detonation at sufficiently low density because of the
rapid increase of $\tau_{\mathrm{nuc}}$ as the bulk expansion causes the density to decrease.
The mechanism of a DDT, however, is likely to require velocity fluctuations that
reach a fraction $\sim 0.1$ of the speed of sound \citep{LisHille00,Woosley07}. A typical figure for the minimal velocity fluctuation $v_{\mathrm{min}}^{\prime}$ on length scales $\ell\sim\ell_{\mathrm{K/RT}}\sim 10\,\mathrm{km}$ is $500\,\mathrm{km\,s^{-1}}$ \citep{WoosPriv}.

Combining the constraints~(\ref{eq:eps_min}) and~(\ref{eq:eps_max}) with the
requirement $\epsilon_{\mathrm{K/RT}}>\epsilon_{\mathrm{min}}:=(v_{\mathrm{min}}^{\prime})^{3}/\ell_{\mathrm{K/RT}}$, the local probability of a DDT becomes
\begin{equation}
\label{eq:prob_eps_da}
\begin{split}
  P\left(\epsilon_{\mathrm{WSR}}>\epsilon_{\ell_{\mathrm{K/RT}}} > \epsilon_{\mathrm{crit}}\right) = &
  \frac{1}{2}\mathrm{erfc}\left[\frac{\ln\left(\max(\epsilon_{\mathrm{crit}},\epsilon_{\mathrm{min}})/\epsilon\right)}{\sqrt{2}\sigma_{\ell_{\mathrm{K/RT}}}}+\frac{\sigma_{\ell_{\mathrm{K/RT}}}}{2\sqrt{2}}\right] \\
  & -
  \frac{1}{2}\mathrm{erfc}\left[\frac{\ln\left(\max(\epsilon_{\mathrm{WSR}},\epsilon_{\mathrm{min}})/\epsilon\right)}{\sqrt{2}\sigma_{\ell_{\mathrm{K/RT}}}}+\frac{\sigma_{\ell_{\mathrm{K/RT}}}}{2\sqrt{2}}\right],
\end{split}
\end{equation}
where $\sigma_{\ell_{\mathrm{K/RT}}}=\sqrt{\mu\ln\left(L/\ell_{\mathrm{K/RT}}\right)}$.
For the asymptotic pdf~(\ref{eq:roepke}) proposed by \citet{Roep07}, on the other hand, it follows that
\begin{equation}
\label{eq:prob_eps_da_roep}
\begin{split}
  P& \left(\epsilon_{\mathrm{WSR}}>\epsilon_{\ell_{\mathrm{K/RT}}} > \epsilon_{\mathrm{crit}}\right) = \frac{\exp(a_{2})}{a_{1}(-a_{0})^{1/a_{1}}}\\ 
&  \times\left\{
\Gamma\left[\frac{1}{a_{1}},-a_{0}\left(\max(\epsilon_{\mathrm{WSR}},\epsilon_{\mathrm{min}})\ell\right)^{a_{1}/3}\right] -
\Gamma\left[\frac{1}{a_{1}},-a_{0}\left(\max(\epsilon_{\mathrm{crit}},\epsilon_{\mathrm{min}})\ell\right)^{a_{1}/3}\right]\right\}.
\end{split}
\end{equation}
The function $\Gamma(x,y)$ is the incomplete gamma function.

As explained in Section~\ref{sc:DDTlam}, the effective DDT probability depending on $\mathrm{Da}_{\,\mathrm{crit}}$ is 
\begin{equation}
\label{eq:prob_ddt_da_tot}
P_{\mathrm{DDT}}(\mathrm{Da}_{\,\mathrm{crit}})=\int_{0}^{\infty}P(\epsilon_{\mathrm{WSR}}>\epsilon_{\ell_{\mathrm{K/RT}}} > \epsilon_{\mathrm{crit}})\,\mathrm{pdf}(\rho|G=0)\,\dd\rho.
\end{equation}
The value of $\mathrm{Da}_{\,\mathrm{crit}}$ has to be determined from microphysical studies. A reasonable range is $10\lesssim\mathrm{Da}_{\,\mathrm{crit}}\lesssim 100$ \citep{WoosPriv}.
Assuming that a DDT occurs at a transition density near $1\times 10^{7}\,\mathrm{g\,cm^{-3}}$ in a region of size $10\,\mathrm{km}$  \citep{WoosKer09}, $\epsilon_{\mathrm{crit}}\sim 2.3\times 10^{19}\,\mathrm{cm^{3}\,s^{-2}}/\mathrm{Da}_{\,\mathrm{crit}}^{3}$. This is, even for $\mathrm{Da}_{\,\mathrm{crit}}\sim 100$, several orders of magnitude greater than $\epsilon_{\mathrm{fl}}\sim 1.1\times 10^{10}\,\mathrm{cm^{3}\,s^{-2}}$. Also note that
$\epsilon_{\mathrm{min}}\sim 10^{17}\,\mathrm{cm^{3}\,s^{-2}}$ is higher than $\epsilon_{\mathrm{crit}}$ for $\mathrm{Da}_{\,\mathrm{crit}}=10$.
For this reason, the probability given by equation~(\ref{eq:prob_ddt_da_tot}) is substantially more constrained than the probability $P_{\mathrm{DDT}}(\mathrm{Ka})$ defined by~(\ref{eq:prob_ddt_ka_tot}). Detailed calculations are presented in the following Section.

\section{Number of deflagration-to-detonation transition events}
\label{sc:number_ddt}

In the previous Section, we obtained estimates of the effective probability of a DDT anywhere close
to the flame surface. In order to assess whether a DDT is likely to occur at some instant,
the expectation value $N_{\mathrm{DDT}}$ of the total number of events has to be calculated. Basically, this means that the effective DDT probability has to be multiplied with the number of regions of a certain size that can be accommodated within the burning zone. In this regard, it is
important to account for the actual flame geometry,  which becomes extremely folded and wrinkled in
the course of the deflagration phase.

Specifying the DDT probability as a function of the Karlovitz number (see formula~\ref{eq:prob_ddt_ka_tot}), we face the difficulty that the size $\ell_{\mathrm{c}}$ of a
region, in which a detonation is triggered by the Zel'dovich mechanism, strongly varies with
the mass density. For this reason, there is no simple relation between $N_{\mathrm{DDT}}$
and $P_{\mathrm{DDT}}(\mathrm{Ka})$. In order to calculate $N_{\mathrm{DDT}}$,
it would be necessary to weigh each $\dd \rho$-bin of $\mathrm{pdf}(\rho|G=0)$ with the
number of regions of size $\ell_{\mathrm{c}}(\rho)$ contained in the volume of burning material
at densities from $\rho$ to $\rho+\dd\rho$. Since $\ell_{\mathrm{c}}$ is much
smaller than the numerical resolution $\Delta$ for $\rho\gtrsim 10^{7}\,\mathrm{g\,cm^{-3}}$,
this weighted distribution cannot be inferred from the numerical data. Nevertheless, 
we can account for the flame geometry as far as it is numerically accessible by
extrapolating the number of grid cells enveloping the flame front, $N_{\Delta}$,
to the number $N_{\mathrm{c}}$ of critical regions of size $\bar{\ell}_{\mathrm{c}}$ given by the mean density of the burning material. If the fractal dimension of the flame front is $D$, it follows that
$N_{\mathrm{c}}\sim N_{\Delta}(\Delta/\bar{\ell}_{\mathrm{c}})^{D}$, and, thus, we estimate the total
number of DDTs to be $N_{\mathrm{DDT}}\sim N_{\mathrm{c}}P_{\mathrm{DDT}}(\mathrm{Ka})$. Of course, the variation of the mass density over the flame surface at a given instant implies that $N_{\mathrm{c}}$ might be substantially different from $N_{\Delta}(\Delta/\bar{\ell}_{\mathrm{c}})^{D}$ even for $\bar{\ell}_{\mathrm{c}}\sim \Delta$. Nevertheless, once $N_{\mathrm{c}}P_{\mathrm{DDT}}(\mathrm{Ka})$ exceeds unity by a great margin, a DDT is likely to occur. Since the flame surface in type Ia supernovae is extremely wrinkled on length scales much smaller than
the integral scale $L$, we tentatively set $D=3$, i.~e., the flame front is assumed to be space-filling.

For the DDT criterion of  \citet{WoosKer09}, on the other hand, an estimate of $N_{\mathrm{DDT}}$
is rather straightforward, because it is assumed that the size of regions, in which a DDT can be triggered, is given by $\ell_{\mathrm{K/RT}}\sim 10\,\mathrm{km}$. This length scale is independent of the mass density, and it is comparable to the size of the grid cells during the pase of the explosion we are interested in. Thus, we have $N_{\mathrm{DDT}}\sim N_{\mathrm{\mathrm{K/RT}}}P_{\mathrm{DDT}}(\mathrm{Da}_{\,\mathrm{crit}})$, where $N_{\mathrm{K/RT}}\sim N_{\Delta}(\Delta/\ell_{\mathrm{K/RT}})^{D}$ and $P_{\mathrm{DDT}}(\mathrm{Da}_{\,\mathrm{crit}})$ is given by equation~(\ref{eq:prob_ddt_da_tot}).

\section{Numerical results}
\label{sc:results}

We interpolated $\epsilon_{\mathrm{fl}}$ from the numerical values of the laminar flame speed and the flame width (defined by the temperature profile) listed in Table~1 of \citet{WoosKer09} as a function of the mass density. Table~3 in the same article specifies values of the nuclear time scale $\tau_{\mathrm{nuc}}$ in the WSR regime for various densities.
To estimate the critical length scale $\ell_{\mathrm{c}}$, we made use of the
compilation of data in Table~1 of \citet{Pan}. Since there are only few values for different mass densities, the numerical evaluation of $\ell_{\mathrm{c}}$ is rather uncertain.

\begin{figure}
\includegraphics[scale=2.0]{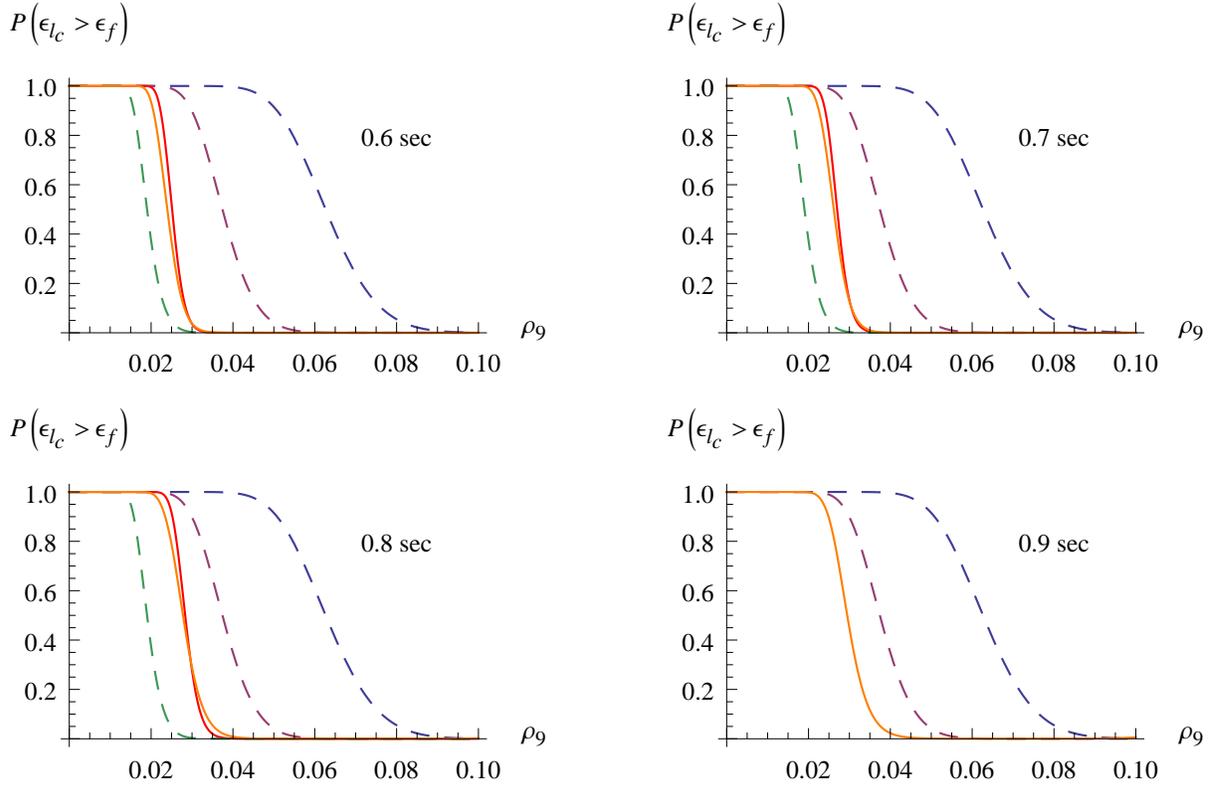}
\caption{Dependence of the probability that the rate of dissipation $\epsilon_{\ell_{\mathrm{c}}}$ in a region of critical size $\ell_{\mathrm{c}}$ exceeds the threshold $\epsilon_{\mathrm{fl}}$ on the mass density for different stages of the explosion. For each instant, the results inferred from angular and radial structure functions (solid lines) are plotted together with three different models (A, C and E) of \citet{Pan}. The mass density is specified in units of $10^{9}\,\mathrm{g\,cm^{-3}}$.}
\label{fig:EpsProb}
\end{figure}

\begin{figure}
\begin{center}
\includegraphics[scale=0.75]{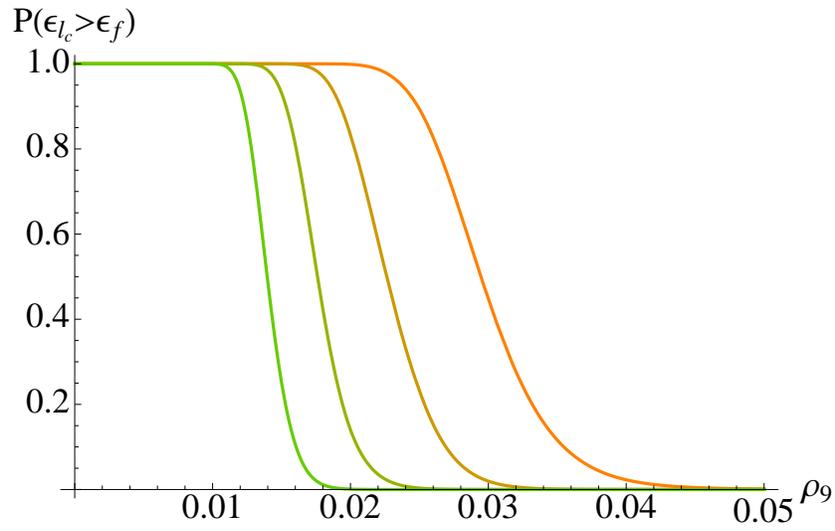}
\end{center}
\caption{Variation of $P\left(\epsilon_{\ell_{\mathrm{c}}} > \mathrm{Ka}^2\epsilon_{\mathrm{fl}}\right)$ with the Karlovitz number $\mathrm{Ka}$ for $t=0.9$ seconds. From right to left, the curves correspond to
$\mathrm{Ka}^{2}=1,\,10,\,100$ and $1000$. The mass density is specified in
units of $10^{9}\,\mathrm{g\,cm^{-3}}$.}
\label{fig:EpsProbKa}
\end{figure}

The resulting probability $P(\epsilon_{\ell_{\mathrm{c}}} > \epsilon_{\mathrm{fl}})$ as function of the mass density for $\mathrm{Ka}=1$ is shown in the Figure~\ref{fig:EpsProb} for $t=0.6$, $0.7$, $0.8$ and $0.9$ seconds. Except for the latest instant of time, the two solid curves
were obtained by substitution of the parameters $\mu$ and $\epsilon$ inferred from
the angular structure functions and the radial structure functions
(see Table~\ref{parameters}).  The differences between both
cases are small. This supports our assumption that the log-normal distribution can be determined
on the basis of the the parameters $\mu_{\mathrm{ang}}$ and $\epsilon_{\mathrm{ang}}$ as well
as $\mu_{\mathrm{rad}}$ and $\epsilon_{\mathrm{rad}}$.
This is important for $t=0.9$ seconds, where the regime of isotropic
turbulence ($\ell<\ell_{\mathrm{K/RT}}$) is numerically unresolved, and we have only
the parameters of the angular structure functions available. 
We find that $P(\epsilon_{\ell_{\mathrm{c}}} > \epsilon_{\mathrm{fl}})\simeq 1$
for $\rho$ less than about $2\times 10^{7}\,\mathrm{g\,cm^{-3}}$
The probabilities assuming the time-independent models A, C and E 
(corresponding to $\epsilon_{\mathrm{fl}}= 10^{16}$, $10^{14}$ and
$10^{12}\,\mathrm{cm^{2}\,s^{-3}}$, respectively) of \citet{Pan} 
are indicated by the dashed lines in Figure~\ref{fig:EpsProb}.
Our data fall in between models C and E. The steeper drop of the
probability toward higher mass density in comparison to the models
of Pan et al. is a consequence of the calculation of $\epsilon_{\mathrm{fl}}$
using the recent data by \citet{WoosKer09}.
Since it is considered to be more likely that a DDT occurs once the Karlovitz number becomes greater than $10$ \citep{WoosKer09}, we also evaluated equation~(\ref{eq:prob_ddt_ka_tot}) for
$\mathrm{Ka}^{2}=10$, $100$ and $1000$. The resulting probabilities
$P(\epsilon_{\ell_{\mathrm{c}}} >\mathrm{Ka}^{2}\epsilon_{\mathrm{fl}})$
are plotted as functions of the mass density in Figure~\ref{fig:EpsProbKa} for $t=0.9$ seconds. As one see, the range of densities for which $P(\epsilon_{\ell_{\mathrm{c}}} >\mathrm{Ka}^{2}\epsilon_{\mathrm{fl}})$ is about unity decreases roughly by a factor of $2$ as $\mathrm{Ka}^{2}$ is raised from $1$ to $1000$.

\begin{figure}
\includegraphics[scale=2.0]{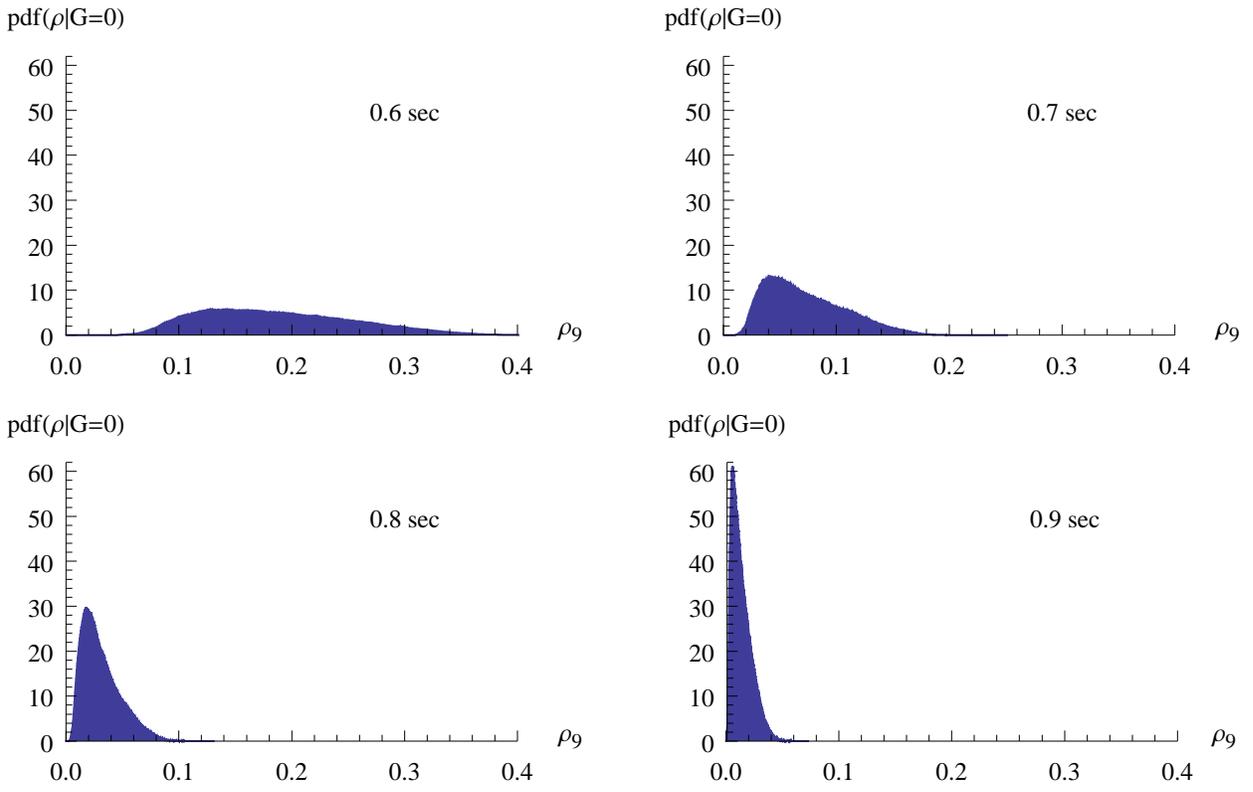}
\caption{Probability density functions of the mass density constrained to the flame front
at different instants. The mass density is specified in units of $10^{9}\,\mathrm{g\,cm^{-3}}$.}
\label{fig:PDF}
\end{figure}

\begin{figure}[t]
\includegraphics[scale=2.0]{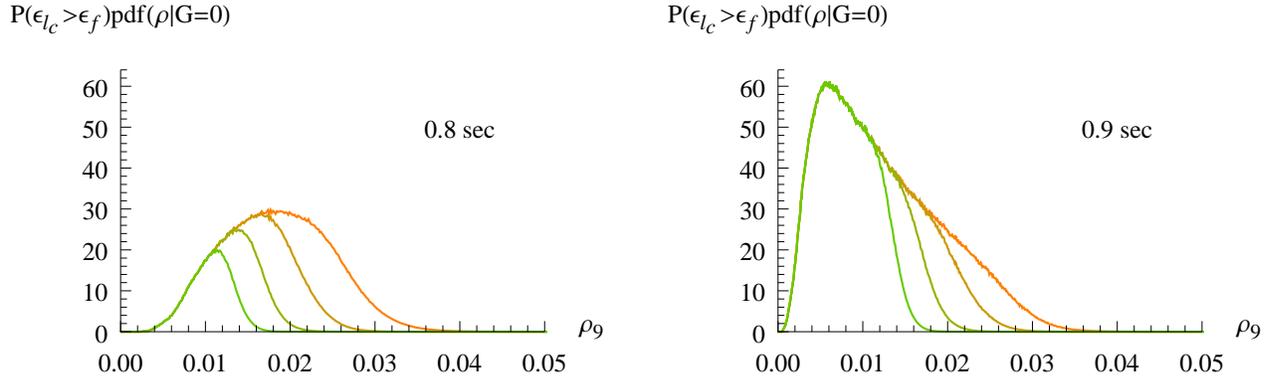}
\caption{Constrained probability density functions for triggering a DDT by velocity fluctuations satisfying
        $\epsilon_{\ell_{\mathrm{c}}} > \mathrm{Ka}^{2}\mathrm\epsilon_{\mathrm{fl}}$ for $t=0.8$ (left) and $0.9$ (right) seconds.
        For $\mathrm{Ka}^{2}$ increasing from $1$ to $1000$ by factors of $10$, the distribution
        becomes increasingly narrow. The calculation is based on the parameters $\mu$ and $\epsilon$ obtained from the angular structure functions. The mass density is specified in
units of $10^{9}\,\mathrm{g\,cm^{-3}}$.}
\label{fig:pdfDDTKa}
\end{figure}

For the computation of the effective probability $P_{\mathrm{DDT}}(\mathrm{Ka})$ as
defined by equation~(\ref{eq:prob_ddt_ka}), we substituted the probability
density functions of the mass density constrained to the 
flame surface, $\mathrm{pdf}(\rho|G=0)$, which are plotted in
Figure~\ref{fig:PDF}.  Comparing Figures~\ref{fig:EpsProb} and
\ref{fig:PDF}, it is evident that $\mathrm{pdf}(\rho|G=0)$
significantly overlaps with $P(\epsilon_{\ell_{\mathrm{c}}} >
\epsilon_{\mathrm{fl}})$ as a function of density for $t\ge 0.8$
seconds only. The product $P(\epsilon_{\ell_{\mathrm{c}}} >
\epsilon_{\mathrm{fl}})\,\mathrm{pdf}(\rho|G=0)$ is shown in
Figure~\ref{fig:pdfDDTKa} for $t=0.8$ and $0.9$ seconds, and the
effective probabilities obtained by integrating the probability densities
are listed in Table~\ref{prob_ln}.  At time earlier than $0.8$
seconds, the resulting probability of a DDT is low.  For
$t=0.8$ seconds, $P_{\mathrm{DDT}}$ would exceed
$50\,\%$, if a DDT was triggered right at the onset of
distributed burning ($\mathrm{Ka}$=1). We emphasize that
this is a highly unrealistic assumption. However, effective
probabilities higher than $50\,\%$ result for all investigated
values of $\mathrm{Ka}$ at time $t=0.9$ seconds.

Because of the huge number of regions of size $\ell_{\mathrm{c}}$ that can be accommodated within the flames, an order-of-magnitude estimate of
$N_{\mathrm{c}}$ as outlined in Section~\ref{sc:number_ddt} implies that $N_{\mathrm{DDT}}$ exceeds unity by many orders of magnitude already at $t=0.6$ seconds. 
At this time, the mass density at the flame front is of the order $10^{8}\,\mathrm{g\,cm^{-3}}$ (see Fig.~\ref{fig:PDF}), and $\ell_{\mathrm{c}}\sim 10^{-3}\,\mathrm{km}$ for this density. The number of grid cells of size $\Delta\approx 4\mathrm{km}$ is $N_{\Delta}\sim 10^{6}$. Hence,
$N_{\mathrm{c}}\sim 10^{17}$ for $D=3$, and with $P_{\mathrm{DDT}}(\mathrm{Ka}=1)\sim 10^{-7}$ (see Table~\ref{prob_ln}), it follows that $N_{\mathrm{DDT}}\sim 10^{10}$. For $\mathrm{Ka}=10$, $P_{\mathrm{DDT}}(\mathrm{Ka})$ decreases by a few orders of magnitude, but $N_{\mathrm{DDT}}$ is still much greater than unity. Even for $D=2$, $N_{\mathrm{c}}\sim 10^{13}$ and $N_{\mathrm{DDT}}\gg 1$ for $\mathrm{Ka}\lesssim10$. Thus, if the conditions for a DDT would be solely constrained by
the Karlovitz number, delayed detonations could easily occur at early stages of the explosion.

\begin{figure}
\begin{center}
\includegraphics[scale=0.75]{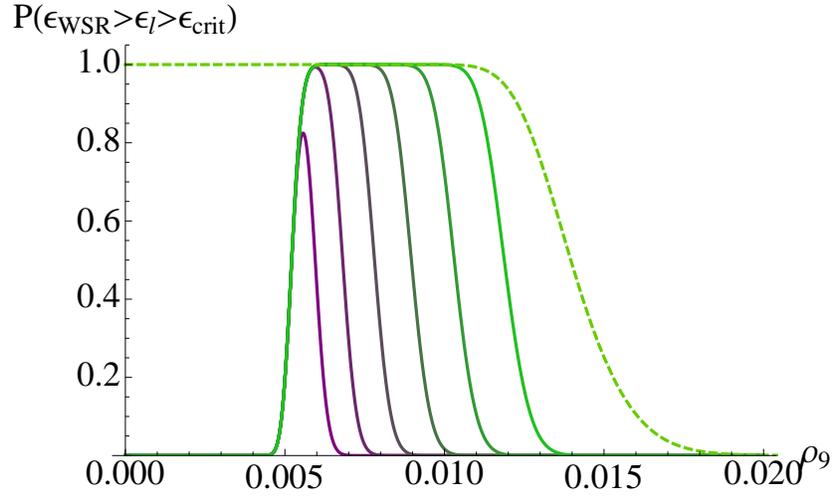}
\end{center}
\caption{Variation of $P\left(\epsilon_{\mathrm{WSR}} > \epsilon_{\ell} > \epsilon_{\mathrm{crit}}\right)$ with the critical Damk\"{o}hler number, $\mathrm{Da}_{\,\mathrm{crit}}$, for $\ell=10\,{\mathrm{km}}$ at time
$t=0.9$ seconds. The distribution becomes increasingly narrow as $\mathrm{Da}_{\,\mathrm{crit}}^3$ decreases
from $10^6$ to $10$ in order-of-magnitude steps. For comparison, $P\left(\epsilon_{\ell_{\mathrm{c}}} > 1000\epsilon_{\mathrm{fl}}\right)$ is shown as dashed curve. The mass density is specified in
units of $10^{9}\,\mathrm{g\,cm^{-3}}$.}
\label{fig:EpsProbDa}
\end{figure}

The graphs of the local probability $P(\epsilon_{\mathrm{WSR}}>\epsilon_{\ell_{\mathrm{K/RT}}} > \epsilon_{\mathrm{crit}})$ according to the DDT constraints proposed by \citet{WoosKer09} are shown in Figure~\ref{fig:EpsProbDa} for $t=0.8$ and $0.9$ seconds. For comparison, $P(\epsilon_{\ell_{\mathrm{c}}} >\mathrm{Ka}^{2}\epsilon_{\mathrm{fl}})$, where $\mathrm{Ka}=10^{3/2}$, is also shown. As $\mathrm{Da}_{\,\mathrm{crit}}$ decreases, the range of densities for which $P(\epsilon_{\mathrm{WSR}}>\epsilon_{\ell_{\mathrm{K/RT}}} > \epsilon_{\mathrm{crit}})\sim 1$ becomes increasingly narrow. In contrast to $P(\epsilon_{\ell_{\mathrm{c}}} >\mathrm{Ka}^{2}\epsilon_{\mathrm{fl}})$, there is also a cutoff toward lower densities, which accounts for the fact that
burning and, consequently, a DDT cannot occur at arbitrarily low densities. However, the
shape of this cutoff is largely qualitative, because it is based on extrapolations
of the microphysical parameters in the range of densities lower than $0.6\times 10^{7}\,\mathrm{g\,cm^{-3}}$. 

\begin{figure}
\begin{center}
\includegraphics[scale=2.0]{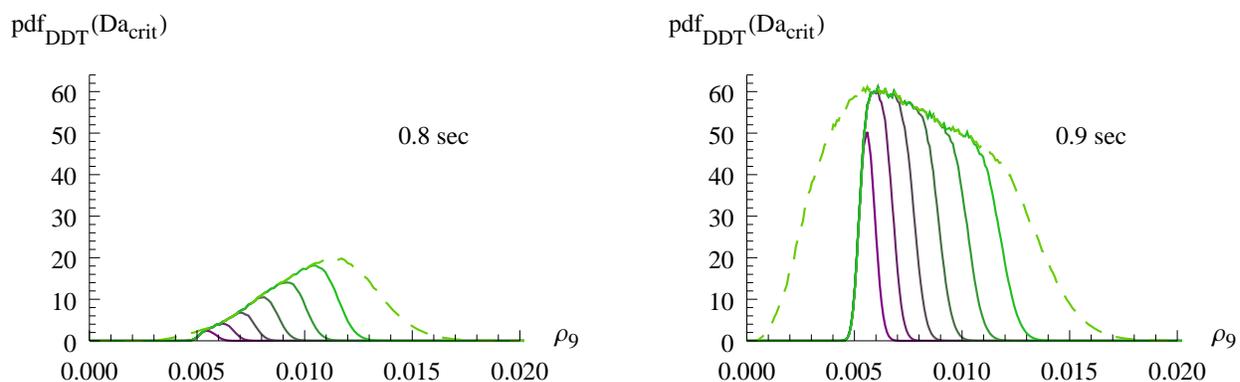}
\end{center}
\caption{Constrained probability density functions for triggering a DDT by velocity fluctuations satisfying
        $\epsilon_{\mathrm{WSR}} > \epsilon_{\ell} > \epsilon_{\mathrm{crit}}$ for $\ell=10\,{\mathrm{km}}$ at time $t=0.8$ (left) and $0.9$ (right) seconds, where the solid curves correspond to
	$\mathrm{Da}_{\,\mathrm{crit}}^3$ decreasing
from $10^6$ to $10$ in order-of-magnitude steps from right to left. The constrained PDF resulting
	from the criterion $\epsilon_{\ell_{\mathrm{c}}} > 1000\epsilon_{\mathrm{fl}}$ is indicated by the dashed line. The calculation is based on the parameters $\mu$ and $\epsilon$ obtained from the angular structure functions. The mass density is specified in units of $10^{9}\,\mathrm{g\,cm^{-3}}$.}
\label{fig:pdfDDTDa}
\end{figure}

Calculating the DDT probability from equation~(\ref{eq:prob_ddt_da_tot}) for critical Damk\"ohler numbers in the range from $10^{1/3}$ to $100$, results in very small probabilities for $t\le 0.7$ seconds (see Table~\ref{prob_da}). From $N_{\mathrm{DDT}}\sim N_{\mathrm{\mathrm{K/RT}}}P_{\mathrm{DDT}}(\mathrm{Da}_{\,\mathrm{crit}})$, it follows that the expectation value of $N_{\mathrm{DDT}}$ is less than unity, except for $\mathrm{Da}_{\,\mathrm{crit}}\ge 100$. For $t=0.8$ and
$0.9$ seconds, on the other hand, the values of $P_{\mathrm{DDT}}(\mathrm{Da}_{\,\mathrm{crit}})$ are much higher and $N_{\mathrm{DDT}}\gg 1$ for all critical Damk\"ohler numbers
if we assume that there is no bound on the minimal velocity fluctuation (i.~e., $\epsilon_{\mathrm{min}}=0$). Consequently, a delayed detonation would occur almost certainly between $0.7$ and $0.8$ seconds after ignition. The constrained probability density functions $P(\epsilon_{\mathrm{WSR}}>\epsilon_{\ell_{\mathrm{K/RT}}} > \epsilon_{\mathrm{crit}})\,\mathrm{pdf}(\rho|G=0)$
plotted in Figure~\ref{fig:pdfDDTDa} allow us to assess the possible range
of transition densities subject to the condition $N_{\mathrm{DDT}}\gtrsim 1$. Depending on the
value of $\mathrm{Da}_{\,\mathrm{crit}}$, transition densities ranging from $0.5\times 10^{7}$ to $1.2\times 10^{7}\,\mathrm{g\,cm^{-3}}$ for $t=0.8$ seconds are most likely. For $\mathrm{Da}_{\,\mathrm{crit}}=10$ a transition density around $0.7\times 10^{7}\,\mathrm{g\,cm^{-3}}$ would be preferred. Since $N_{\mathrm{DDT}}\sim 10^{4}$, the transition density might be closer to $10^{7}\,\mathrm{g\,cm^{-3}}$ though.
It is reassuring to note that our DDT densities are comparable to the ones determined from fits of 1D delayed detonation models to observational data. However, an exact match is not expected and would not necessarily lead to agreement of multi-dimensional models with observations because the detonation wave does not propagate spherically outwards.

\begin{figure}
\begin{center}
\includegraphics[scale=0.75]{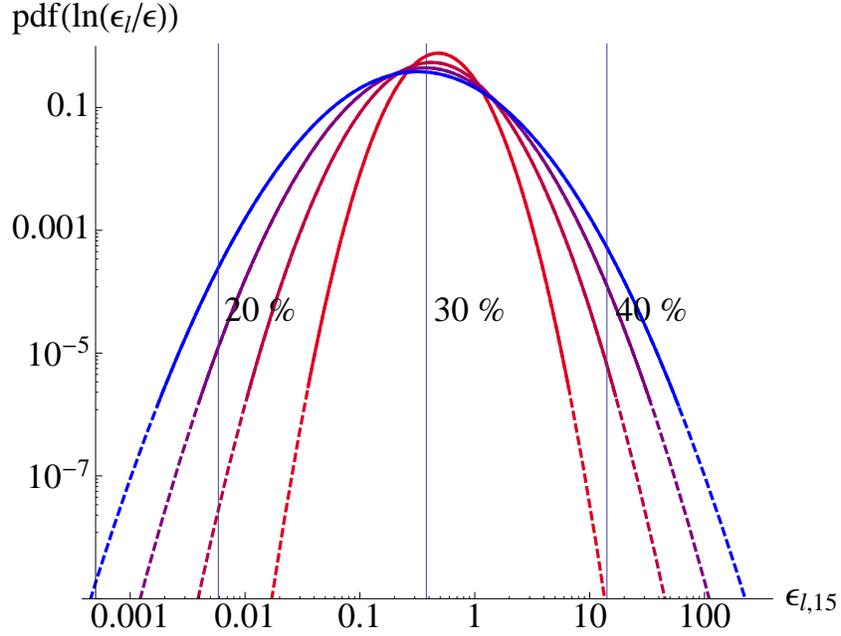}
\end{center}
        \caption{Log-normal probability density functions~(\ref{eq:lognorm}) of 
        $\epsilon_{\mathrm{\ell}}$ in units of $10^{15}\,\mathrm{cm^{3}\,s^{-2}}$ for 
        $\ell = 10\,\mathrm{km}$ at time $t=0.9$ seconds. The mean rate of dissipation 
        $\epsilon$ and the integral scale $L$ are determined by the third-order angular structure
        functions (see Table~\ref{parameters}). The four plotted pdfs follow for $mu=0.05$ 
        (most narrow distribution), $0.1$, $0.15$, and $0.2$ (widest distribution).        
        The tails beyond $5\sigma_{\ell}$ are dashed. The vertical lines indicate the values of 
        $\epsilon_{\mathrm{crit}}$ defined by equation~(\ref{eq:eps_min}) for different mass densities,
        assuming that the critical Damk\"{o}hler number is equal to $10$. The percentages indicate the
        volume fractions of matter at densities less than the mass densities chosen for the evaluation of 
        $\epsilon_{\mathrm{crit}}$.}
         \label{fig:lognorm}
\end{figure}

However, the probability of a DDT drops dramatically if the minimal velocity to
trigger a detonation, $v_{\mathrm{min}}^{\prime}$, is greater than about $100\,\mathrm{km s^{-1}}$. In Table~\ref{prob_da_mu}, the results for $v_{\mathrm{min}}^{\prime}=0$, $200$ and $500\,\mathrm{km s^{-1}}$ are compared for different values of the intermittency parameter $\mu$ at time $t=0.9$ seconds. As one can see, delayed detonations are virtually excluded if $v_{\mathrm{min}}^{\prime}=500\,\mathrm{km s^{-1}}$, as proposed by \citet{Woosley07}. Assuming a smaller minimal velocity of $200\,\mathrm{km s^{-1}}$, the results
turn out to be extremely sensitive on the intermittency parameter $\mu$, while only small variation is
found for $v_{\mathrm{min}}^{\prime}=0\,\mathrm{km s^{-1}}$. The reason becomes apparent from the
dependence of the log-normal distribution on the intermittency parameter.
In Figure~\ref{fig:lognorm}, log-normal probability density functions~\ref{eq:lognorm} for $\ell=10\,\mathrm{km}$, $\mu=0.05,\ldots,0.2$, and the parameters $L$ and $\epsilon$ determined by the angular structure
functions at time $t=0.9$ seconds are plotted. Assuming $\mathrm{Da}_{\,\mathrm{crit}}=10$, the thresholds $\epsilon_{\mathrm{crit}}$ defined by equation~(\ref{eq:eps_min}) are indicated for several different mass densities, which correspond to cumulative mass fractions of $20$, $30$, and $40\,\%$ of burning material.
If we solely consider the constraint $\epsilon_{\mathrm{WSR}}>\epsilon_{\ell}>\epsilon_{\mathrm{crit}}$, the intermittency parameter has only a small influence, because the cumulative
probability is mainly determined by the central parts of the distributions. In the case that $\epsilon_{\ell}$ must further satisfy, say, $\epsilon_{\ell}>\epsilon_{\min}=8\times 10^{15}\,\mathrm{cm^{3}\,s^{-2}}$ corresponding to $v_{\mathrm{min}}^{\prime}=200\,\mathrm{km s^{-1}}$, the cumulative probabilities are given by the right wings of the log-normal pdfs. Depending on the value of $\mu$, there are substantial variations of the effective probabilities, as different portions of the wings contribute. If $\epsilon_{\min}$ exceeds about $10^{17}\,\mathrm{cm^{3}\,s^{-2}}$, the probability of a DDT becomes virtually zero. 

\begin{figure}
\begin{center}
\includegraphics[scale=2.0]{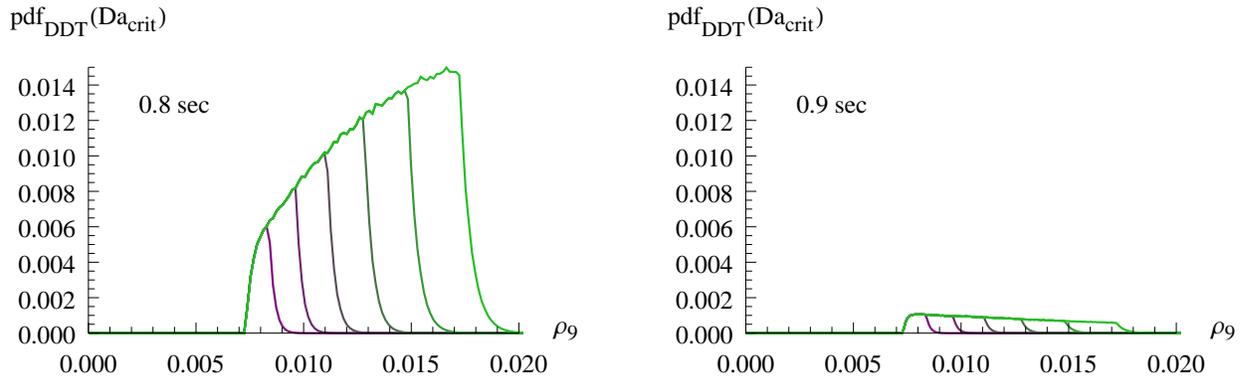}
\end{center}
\caption{The same probability density functions as in Figure~\ref{fig:pdfDDTDa}, however, with the local DDT probability given by
	equation~(\ref{eq:prob_eps_da_roep}) in place of~(\ref{eq:prob_eps_da}), and $v_{\mathrm{min}}^{\prime}=500\,\mathrm{km s^{-1}}$. Note that the ordinate scale is different from
	Figure~\ref{fig:pdfDDTDa}.
\label{fig:pdfDDTDaRoep}}
\end{figure}

\begin{figure}
\begin{center}
\includegraphics[scale=2.0]{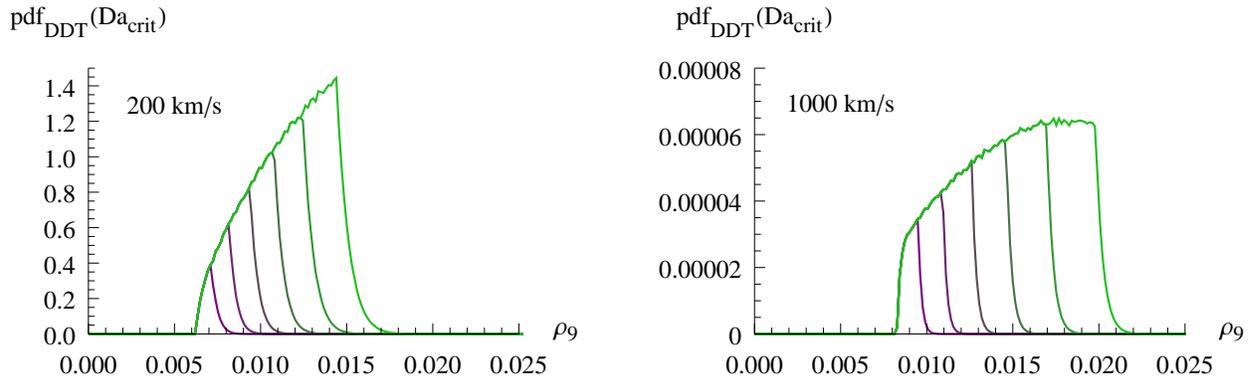}
\end{center}
\caption{Variation of the probability density functions shown in the left panel of Fig.~\ref{fig:pdfDDTDaRoep} for
$v_{\mathrm{min}}^{\prime}=200\,\mathrm{km s^{-1}}$ (left) and $v_{\mathrm{min}}^{\prime}=1000\,\mathrm{km s^{-1}}$ (right). Note that the different ordinate scales.
\label{fig:pdfDDTDaRoep_vmin}}
\end{figure}

The asymptotic tails of the $\epsilon_{\ell}$-pdfs calculated by \citet{Roep07}, on the other hand, permit delayed detonations even for $v_{\mathrm{min}}^{\prime}=500\,\mathrm{km s^{-1}}$. Table~\ref{prob_da_roep} summerizes the
results for $P_{\mathrm{DDT}}$ and $N_{\mathrm{DDT}}$ at $t=0.7$, $0.8$ and $0.9$ seconds. For $v_{\mathrm{min}}^{\prime}=500\,\mathrm{km s^{-1}}$, DDTs are likely to occur between $0.7$ and $0.8$ seconds. The constrained probability density function at $t=0.8$ seconds is plotted in the left panel of  Fig.~\ref{fig:pdfDDTDaRoep}. The transition density for $\mathrm{Da}_{\,\mathrm{crit}}=10$ is close to $1.0\times 10^{7}\,\mathrm{g\,cm^{-3}}$. After $0.8$ seconds the probabilities begin to decline because of the
decreasing turbulence intensity (see Table~\ref{prob_da_roep} and the right panel of 
Fig.~\ref{fig:pdfDDTDaRoep}).
This behavior is in marked contrast to the predictions based on lognormal distributions,
for which the DDT probabilities at $t=0.9$ seconds are higher than for $0.8$ seconds
(see Table~\ref{prob_da} and Fig.~\ref{fig:pdfDDTDa}). Therefore, there appears to be a narrow time window,
in which DDTs are possible. In Fig.~\ref{fig:pdfDDTDaRoep_vmin}, plots of $\mathrm{pdf}_{\mathrm{DDT}}(\mathrm{Da}_{\,\mathrm{crit}})$ for $t=0.8$ seconds are shown for
$v_{\mathrm{min}}^{\prime}=200$ and $1000\,\mathrm{km s^{-1}}$. It is obvious that the value of
$v_{\mathrm{min}}^{\prime}$ has a huge impact on the probability density functions. As a consequence,
delayed detonations could be set off earlier than $0.7$ seconds after ignition if $v_{\mathrm{min}}$ was
about $200\,\mathrm{km s^{-1}}$ or less, whereas the DDT probabilities would be marginal for $v_{\mathrm{min}}^{\prime}=1000\,\mathrm{km s^{-1}}$.

\section{Conclusion}

We investigated the intermittency properties of turbulence in the
numerical simulation of a thermonuclear supernova by
\citet{Roepke}. \citet{Pan} proposed that the probability of
entering the distributed burning regime can be computed from the
log-normal model for intermittent turbulence, and from this
probability the incidence of a DDT can be 
inferred. Evaluating the characteristic
scales of turbulence and the probability density functions of the mass
density in the vicinity of the flame front at different instants, we calculated the
probability of a DDT for various Karlovitz numbers. 
We also investigated the influence of more
restrictive criteria following from the numerical studies by
\cite{WoosKer09}.

Assuming that a detonation can be triggered after the onset of distributed burning,
our calculations indicate that a delayed detonation would commence 
at an early stage of the explosion. In contrast to \citet{Pan}, where a spherical flame is assumed, 
the highly wrinkled and folded flame front in the numerical simulation
greatly increases the number of regions of critical size. Although
we cannot precisely determine the total number of regions of size $\ell_{\mathrm{c}}$,
because the critical size greatly varies with the mass density \citep{Niemeyer2},
and $\ell_{\mathrm{c}}\ll\Delta$, where $\Delta$ is the grid resolution,
it appears that $N_{\mathrm{DDT}}$ greatly exceeds unity already $0.6$ seconds
after ignition. Then the transition density would be signifcantly higher than 
$10^{7}\,\mathrm{g\,cm^{-3}}$. This possibility was pointed out by \citet{Pan}. 

If DDT conditions are constrained by an interval of Damk\"{o}hler numbers in
the stirred flame regime \citep{WoosKer09}, a DDT earlier than $0.7$ seconds
after ignition can be excluded. Assuming a typical value of the Damk\"{o}hler number,
for which a DDT can be triggered, a delayed detonation could occur around $t=0.8$ seconds,
and the transition density would be close to $10^{7}\,\mathrm{g\,cm^{-3}}$. However,
assuming a log-normal distribution, the typcial velocity fluctuations would be
implausibly low. If turbulent velocity fluctuations greater than $500\,\mathrm{km s^{-1}}$
are required \citep{WoosPriv}, then non-log-normal distributions are vital for
delayed detonations.

One of the problems of determining log-normal distributions from scaling exponents
of structure functions is that it is difficult to calculate higher-order two-point
statistics. Consequently, the computed scaling exponents are sufficient to 
constrain the central part of the distribution only, whereas the tails are
based on an extrapolation. A further caveat is that we are not able to separate ash from fuel regions,
because two-point statistics of turbulence can only be computed in convex regions. 
For DDTs, however, turbulence in fuel close to the flame
front is signicant.
Indeed, \citet{Roep07} showed that the tail of the pdf of turbulent velocity fluctuations
on a length scale of about $10\,\mathrm{km}$ in the vincity of the flame front is much
flatter than what is expected on the basis of log-normal models. With this distribution,
DDTs between $0.7$ and $0.8$ seconds after ignition at a transition density close
to $10^{7}\,\mathrm{g\,cm^{-3}}$ are definitely possible. Of course, the detonation
time also depends on the ignition scenario and the subsequent evolution of the
deflagration \citep[see, for example,][]{SchmNie06}. In any case, it is crucial
to settle the question of the relevant distribution of turbulent velocity fluctuations in
the future. Given a certain distribution, the minimal velocity fluctuation
that is necessary to trigger a DDT is an extremely important parameter. Depending
on this threshold, delayed detonations might be theoretically confirmed or excluded.
Consequently, further studies of the microphysics, particularly in the range of densities below $10^{7}\,\mathrm{g\,cm^{-3}}$, will be essential
for a more accurate evaluation of the DDT probability in type Ia supernovae. 

Based on the insights provided by such calculations, it may be possible to devise
an algorithm that determines the local probability of a DDT in
large-scale simulations of thermonunclear supernovae. Since the
processes causing a DDT occur on scales that are difficult to resolve in such simulations, a subgrid scale model is likely to play some role \citep{SchmNie06b}. Apart from this, the propagation of the burning zone has to be treated beyond the onset of distributed burning. 
\citet{Schm07} proposed that the level set technique
can be extended at least into the broken-reaction-zones regime \citep{KimMen00}, provided that
the burning time scale does not exceed some fraction of the eddy turn-over time corresponding
to the numerically unresolved velocity fluctuations (i.~e., the ratio of the
grid cell size to the square root of the specific subgrid scale turbulence
energy). This approach also calls for further microphysical studies. 
An entirely different approach might be the use of adaptive
mesh refinement and the \emph{in situ} calculation of the
processes triggering a DDT, for instance, by means of LEM \citep{Kerst91}.
Once all numerical challenges are met, quantitative theoretical arguments in favor or against delayed detonations as an explanation for type Ia supernovae will be within reach.

\acknowledgments

We are grateful to Stanford Woosley and Alan Kerstein for their comments
and suggestions. The research of F.\ K.\ R\"opke is supported through the Emmy Noether Program
of the German Research Foundation (DFG; RO~3676/1-1) and by the
Excellence Cluster ``Origin and Structure of the Universe''.




\clearpage




\clearpage




\clearpage

\clearpage

\clearpage

\clearpage

\clearpage

\clearpage

\clearpage

\clearpage

\clearpage

\clearpage

\clearpage

\begin{table}
  \begin{center}
      \begin{tabular}{lrrllcccccc}
        \tableline\tableline
        $t$ [s] & $L$ & $l_{\mathrm{K/RT}}$ & $\epsilon_{\mathrm{ang}}$ & $\epsilon_{\mathrm{rad}}$ & $\mu_{\mathrm{ang}}$ & $\mu_{\mathrm{rad}}$ &
        $\beta_{\mathrm{ang}}$ & $\beta_{\mathrm{rad}}$ &
        $\Delta_{\mathrm{ang}}$ & $\Delta_{\mathrm{rad}}$\\
        \tableline
        0.5 &  516  & 15.4 & 0.0332 & 0.0634 & 0.096 & 0.139 & 0.783 & 0.738 & 0.434 & 0.523\\
        0.6 &  799  & 14.4 & 0.0819 & 0.0972 & 0.128 & 0.134 & 0.750 & 0.744 & 0.501 & 0.512\\
        0.7 & 1023  & 14.3 & 0.166  & 0.177  & 0.111 & 0.121 & 0.766 & 0.755 & 0.468 & 0.489\\
        0.8 & 1578  & 12.9 & 0.335  & 0.274  & 0.140 & 0.128 & 0.738 & 0.749 & 0.525 & 0.503\\
        0.9 & 2324  & \nodata & 0.551 & \nodata & 0.132 & \nodata & 0.745 & \nodata & 0.510 & \nodata\\
        \tableline
      \end{tabular}
  \end{center}
  \caption{Parameters of log-normal and log-Poisson model fits to
        the angular and radial scaling exponents at different instants. The
        radial scalings apply to the subrange $\ell<\ell_{\mathrm{K/RT}}$.
        Also specified are the integral scale and the mean rate of dissipation
        inferred from the radial and angular structure functions.
  \label{parameters}}
\end{table}

\begin{table}
  \begin{center}
      \begin{tabular}{lcccccc}
        \tableline\tableline
        $t$ [s] & $\mu_{\mathrm{ang}}$ & $2-\zeta_{6,\mathrm{ang}}$ & 
        $\mathrm{std}(\zeta_{6,\mathrm{ang}})$ & $\mu_{\mathrm{rad}}$ & 
        $2-\zeta_{6,\mathrm{rad}}$ & $\mathrm{std}(\zeta_{6,\mathrm{rad}})$ \\
        \tableline
        0.5 & 0.096 & 0.099 & 0.037 & 0.139 & 0.144 & 0.047 \\ 
        0.6 & 0.128 & 0.125 & 0.033 & 0.134 & 0.141 & 0.055 \\
        0.7 & 0.111 & 0.109 & 0.028 & 0.121 & 0.129 & 0.057 \\
        0.8 & 0.140 & 0.141 & 0.044 & 0.128 & 0.134 & 0.054 \\
        0.9 & 0.132 & 0.132 & 0.043 & \nodata & \nodata & \nodata \\
        \tableline
      \end{tabular}
  \end{center}
  \caption{Fitted intermittency parameter of the log-normal model and values following from
the sixth-order scaling exponents of the angular and radial structure functions. Also listed are the standard errors of the scaling exponents following from power-law fits to the structure functions (see paper I).
  \label{zeta_6}}
\end{table}

\begin{table}
  \begin{center}
      \begin{tabular}{cllllll}
        \tableline\tableline
        & \multicolumn{2}{c}{$\mathrm{Ka}=1$} & \multicolumn{2}{c}{$\mathrm{Ka}=10$} & \multicolumn{2}{c}{$\mathrm{Ka}=31.6$}  \\
        $t$ [s] & $P_{\mathrm{DDT}}^{(\mathrm{ang})}$ & $P_{\mathrm{DDT}}^{(\mathrm{rad})}$ & $P_{\mathrm{DDT}}^{(\mathrm{ang})}$ & $P_{\mathrm{DDT}}^{(\mathrm{rad})}$ & $P_{\mathrm{DDT}}^{(\mathrm{ang})}$ & $P_{\mathrm{DDT}}^{(\mathrm{rad})}$\\ 
        \tableline
        0.6 & $3.0\times 10^{-7}$ & $4.8\times 10^{-7}$ & $\sim 0$ & $\sim 0$ & $\sim 0$ & $\sim 0$\\
        0.7 & $0.045$ & $0.050$ & $0.0024$ & $0.0026$ & $4.2\times 10^{-4}$ & $4.8\times 10^{-4}$ \\
        0.8 & $0.52$ & $0.53$ & $0.22$ & $0.22$ & $0.12$ & $0.13$ \\
        0.9 & $0.93$ & \nodata & $0.72$ & \nodata & $0.59$ & \nodata \\ 
        \tableline
      \end{tabular}
  \end{center}
  \caption{Effective probability of a DDT inferred from equation~(\ref{eq:prob_ddt_ka}) for different
  values of the Karlovitz number.
  \label{prob_ln}}
\end{table}

\begin{table}
  \begin{center}
      \begin{tabular}{cllll}
        \tableline\tableline
        $\mathrm{Da}_{\,\mathrm{crit}}$ & 
	$P_{\mathrm{DDT}}^{(\mathrm{ang})}$ & $N_{\mathrm{DDT}}^{(\mathrm{ang})}$ & 
	$P_{\mathrm{DDT}}^{(\mathrm{rad})}$ & $N_{\mathrm{DDT}}^{(\mathrm{rad})}$\\ 
       \tableline                
       \multicolumn{5}{c}{$t=0.7$ seconds, $N_{\mathrm{K/RT}}\approx 6.1\cdot 10^{5}$}\\
       \tableline
       2.15 & $\sim 0$ & $\sim 0$ & $\sim 0$ & $\sim 0$ \\
       4.64 & $\sim 0$ & $\sim 0$ & $\sim 0$ & $\sim 0$ \\
       10.0 & $\sim 0$ & $\sim 0$ & $\sim 0$ & $\sim 0$ \\
       21.5 & $1.1\times 10^{-8}$ & $7.0\times 10^{-3}$ & $\sim 0$ & $\sim 0$ \\
       46.4 & $4.4\times 10^{-6}$ & $2.7$ & $4.1\times 10^{-6}$ & $2.5$\\
       100  & $7.6\times 10^{-5}$ & $4.7\times 10^{1}$ & $8.6\times 10^{-5}$ & $5.3\times 10^{1}$\\
       \tableline                
       \multicolumn{5}{c}{$t=0.8$ seconds, $N_{\mathrm{K/RT}}\approx 1.7\cdot 10^{6}$}\\
        \tableline
       2.15 & 0.0021 & $3.9\times 10^{3}$ & 0.0022 & $3.7\times 10^{3}$\\
       4.64 & 0.0061 & $1.1\times 10^{4}$ & 0.0061 & $1.1\times 10^{4}$\\
       10.0 & 0.013  & $2.4\times 10^{4}$ & 0.014 & $2.4\times 10^{4}$\\
       21.5 & 0.026 & $4.8\times 10^{4}$ & 0.027 & $4.7\times 10^{4}$\\
       46.4 & 0.046 & $8.4\times 10^{4}$ &  0.047 & $8.2\times 10^{4}$\\
       100  & 0.076 & $1.4\times 10^{5}$ & 0.077 & $1.3\times 10^{5}$\\
       \tableline                
       \multicolumn{5}{c}{$t=0.9$ seconds, $N_{\mathrm{K/RT}}\approx 4.1\cdot 10^{6}$}\\
        \tableline
       2.15 & 0.04 & $1.8\times 10^{5}$ & \nodata & \nodata\\
       4.64 & 0.10 & $3.9\times 10^{5}$ & \nodata & \nodata\\
       10.0 & 0.15 & $6.3\times 10^{5}$ & \nodata & \nodata\\
       21.5 & 0.22 & $8.8\times 10^{5}$ & \nodata & \nodata\\ 
       46.4 & 0.28 & $1.2\times 10^{6}$  & \nodata & \nodata\\
       100  & 0.36 & $1.5\times 10^{6}$  & \nodata & \nodata\\
        \tableline
      \end{tabular}
  \end{center}
  \caption{Dependence of the effective probability of a DDT and the expectation value of the number of DDTs
	on the critical Damk\"{o}hler number for several instants.
  \label{prob_da}}
\end{table}

\begin{table}
  \begin{center}
      \begin{tabular}{lllllll}
        \tableline\tableline
	& \multicolumn{2}{c}{$v_{\mathrm{min}}^{\prime}=0\,\mathrm{km\,s^{-1}}$} &
	\multicolumn{2}{c}{$v_{\mathrm{min}}^{\prime}=200\,\mathrm{km\,s^{-1}}$} &
	\multicolumn{2}{c}{$v_{\mathrm{min}}^{\prime}=500\,\mathrm{km\,s^{-1}}$} \\
        $\mu$ & $P_{\mathrm{DDT}}$ & $N_{\mathrm{DDT}}$ &
	$P_{\mathrm{DDT}}$ & $N_{\mathrm{DDT}}$ & $P_{\mathrm{DDT}}$ & $N_{\mathrm{DDT}}$\\
       \tableline
       0.05 & $0.155$ & $ 6.34\times 10^{5}$ & $6.4\times 10^{-9}$ & $2.6\times 10^{-2}$ & $\sim 0$ & $\sim 0$ \\ 
       0.1   & $0.154$ & $6.30\times 10^{5}$ & $5.8\times 10^{-6}$ & $2.4\times 10^{1}$ & $\sim 0$ & $\sim 0$ \\
       0.15 & $0.153$ & $6.25\times 10^{5}$ & $5.8\times 10^{-5}$ & $2.4\times 10^{2}$ & $\sim 0$ & $\sim 0$ \\
       0.2   & $0.152$ & $6.20\times 10^{5}$ & $1.8\times 10^{-4}$ & $7.5\times 10^{2}$ & $1.1\times 10^{-9}$ & $4.4\times 10^{-3}$ \\
         \tableline
      \end{tabular}
  \end{center}
  \caption{Dependence of the effective probability of a DDT and the expectation value of the number of DDTs
	on the intermittency parameter $\mu$ of the log-normal model and the minimal velocity fluctuation
	$v_{\mathrm{min}}^{\prime}$ for $t=0.9$ seconds.
  \label{prob_da_mu}}
\end{table}

\begin{table}
  \begin{center}
      \begin{tabular}{cllllll}
        \tableline\tableline
	& \multicolumn{2}{c}{$v_{\mathrm{min}}^{\prime}=200\,\mathrm{km\,s^{-1}}$} &
	\multicolumn{2}{c}{$v_{\mathrm{min}}^{\prime}=500\,\mathrm{km\,s^{-1}}$} &
	\multicolumn{2}{c}{$v_{\mathrm{min}}^{\prime}=1000\,\mathrm{km\,s^{-1}}$} \\
        $\mathrm{Da}_{\,\mathrm{crit}}$ & $P_{\mathrm{DDT}}$ & $N_{\mathrm{DDT}}$ &
	$P_{\mathrm{DDT}}$ & $N_{\mathrm{DDT}}$ & $P_{\mathrm{DDT}}$ & $N_{\mathrm{DDT}}$\\
       \tableline                
       \multicolumn{7}{c}{$t=0.7$ seconds, $N_{\mathrm{K/RT}}\approx 6.1\cdot 10^{5}$}\\
       \tableline
       2.15 & $\sim 0$            & $\sim 0$            & $\sim 0$            & $\sim 0$            & $\sim 0$ & $\sim 0$ \\
       4.64 & $3.7\times 10^{-9}$ & $2.3\times 10^{-3}$ & $2.3\times 10^{-9}$ & $1.4\times 10^{-3}$ & $\sim 0$ & $\sim 0$ \\
       10.0 & $3.9\times 10^{-7}$ & $2.4\times 10^{-1}$ & $4.2\times 10^{-8}$ & $2.6\times 10^{-2}$ & $\sim 0$ & $\sim 0$ \\
       21.5 & $5.9\times 10^{-6}$ & $3.6$               & $2.1\times 10^{-7}$ & $1.3\times 10^{-1}$ & $\sim 0$ & $\sim 0$ \\
       46.4 & $3.0\times 10^{-5}$ & $1.9\times 10^{1}$  & $6.6\times 10^{-7}$ & $4.0\times 10^{-1}$ & $1.7\times 10^{-9}$ & $1.0\times 10^{-3}$ \\ 
       100  & $9.5\times 10^{-5}$ & $5.8\times 10^{1}$  & $1.8\times 10^{-6}$ & $1.1$               & $4.5\times 10^{-9}$ & $2.8\times 10^{-3}$ \\ 
       \tableline                
       \multicolumn{7}{c}{$t=0.8$ seconds, $N_{\mathrm{K/RT}}\approx 1.7\cdot 10^{6}$}\\
        \tableline
       2.15 & $3.5\times 10^{-4}$ & $6.2\times 10^{2}$ & $6.2\times 10^{-6}$ & $1.1\times 10^{1}$ & $3.9\times 10^{-8}$ & $6.8\times 10^{-2}$ \\
       4.64 & $9.9\times 10^{-4}$ & $1.7\times 10^{3}$ & $1.6\times 10^{-5}$ & $2.8\times 10^{1}$ & $9.8\times 10^{-8}$ & $1.7\times 10^{-1}$ \\
       10.0 & $2.0\times 10^{-3}$ & $3.5\times 10^{3}$ & $3.1\times 10^{-5}$ & $5.3\times 10^{1}$ & $1.8\times 10^{-7}$ & $3.2\times 10^{-1}$ \\
       21.5 & $3.5\times 10^{-3}$ & $6.1\times 10^{3}$ & $5.1\times 10^{-5}$ & $8.9\times 10^{1}$ & $3.0\times 10^{-7}$ & $5.1\times 10^{-1}$ \\
       46.4 & $5.6\times 10^{-3}$ & $9.8\times 10^{3}$ & $7.9\times 10^{-5}$ & $1.4\times 10^{2}$ & $4.5\times 10^{-7}$ & $7.8\times 10^{-1}$ \\
       100  & $8.5\times 10^{-3}$ & $1.5\times 10^{4}$ & $1.2\times 10^{-4}$ & $2.0\times 10^{2}$ & $6.3\times 10^{-7}$ & $1.1$ \\
       \tableline                
       \multicolumn{7}{c}{$t=0.9$ seconds, $N_{\mathrm{K/RT}}\approx 4.1\cdot 10^{6}$}\\
        \tableline
       2.15 & $6.1\times 10^{-4}$ & $2.5\times 10^{3}$ & $1.2\times 10^{-6}$ & $4.8$              & $\sim 0$            & $\sim 0$ \\
       4.64 & $1.3\times 10^{-3}$ & $5.3\times 10^{3}$ & $2.4\times 10^{-6}$ & $9.9$              & $1.2\times 10^{-9}$ & $5.0\times 10^{-3}$ \\
       10.0 & $2.0\times 10^{-3}$ & $8.3\times 10^{3}$ & $3.8\times 10^{-6}$ & $1.6\times 10^{1}$ & $1.9\times 10^{-9}$ & $7.8\times 10^{-3}$ \\
       21.5 & $2.8\times 10^{-3}$ & $1.2\times 10^{4}$ & $5.2\times 10^{-6}$ & $2.1\times 10^{1}$ & $2.6\times 10^{-9}$ & $1.1\times 10^{-2}$ \\
       46.4 & $3.7\times 10^{-3}$ & $1.5\times 10^{4}$ & $6.7\times 10^{-6}$ & $2.8\times 10^{1}$ & $3.3\times 10^{-9}$ & $1.4\times 10^{-2}$ \\
       100  & $4.6\times 10^{-3}$ & $1.9\times 10^{4}$ & $8.3\times 10^{-6}$ & $3.4\times 10^{1}$ & $4.0\times 10^{-9}$ & $1.6\times 10^{-2}$ \\
        \tableline
      \end{tabular}
  \end{center}
  \caption{Dependence of the effective probability of a DDT and the expectation value of the number of DDTs on the critical Damk\"{o}hler number corresponding to Table~\ref{prob_da} for a non-log-normal distribution of turbulent velocity fluctuations \citep{Roep07}.
  \label{prob_da_roep}}
\end{table}


\begin{thebibliography}{19}
\expandafter\ifx\csname natexlab\endcsname\relax\def\natexlab#1{#1}\fi

\bibitem[{{Boldyrev} {et~al.}(2002){Boldyrev}, {Nordlund}, \&
  {Padoan}}]{Boldyrev}
{Boldyrev}, S., {Nordlund}, {\AA}., \& {Padoan}, P. 2002, \apj, 573, 678

\bibitem[{{Ciaraldi-Schoolmann} {et~al.}(2008)}]{CirSchm08}
{Ciaraldi-Schoolmann}, F., {Schmidt}, W., {Niemeyer}, J.~C., {R{\"o}pke}, F.~K.,\& {Hillebrandt}, W. 2008, \apj, in press (arXiv:0901.4254)

\bibitem[{{Dubrulle}(1994)}]{Dubr}
{Dubrulle}, B. 1994, Physical Review Letters, 73, 959

\bibitem[{{Frisch}(1995)}]{Frisch}
{Frisch}, U. 1995, {Turbulence. The legacy of A.N. Kolmogorov} (Cambridge:
  Cambridge University Press, 1995)

\bibitem[{{Gamezo} {et~al.}(1991)}]{Gam03}
{Gamezo}, V.~N., {Khokhlov}, A.~M., {Oran}, E.~S., {Chtchelkanova},
 A.~Y., \& {Rosenberg}, R.~O. 2003, Science, 299, 77

\bibitem[{{Hillebrandt} \& {Niemeyer}(2000)}]{HilleNie00}
{Hillebrandt}, W. \& {Niemeyer}, J.~C. 2000, \araa, 38, 191

\bibitem[{{Kerstein}(1991)}]{Kerst91}
{Kerstein}, A.~R. 1991, J. Fluid Mech., 231, 361

\bibitem[{{Kerstein}(2001)}]{Kerst01}
{Kerstein}, A.~R. 2001, \pre, 64, 066306

\bibitem[{{Khokhlov}(1991)}]{Khokhlov3}
{Khokhlov}, A.~M. 1991, \aap, 245, 114

\bibitem[{{Khokhlov} {et~al.}(1997){Khokhlov}, {Oran}, \&
  {Wheeler}}]{Khokhlov4}
{Khokhlov}, A.~M., {Oran}, E.~S., \& {Wheeler}, J.~C. 1997, \apj, 478, 678

\bibitem[{{Kim} \& {Menon}(2000)}]{KimMen00}
{Kim}, W., \& {Menon}, S. 2000, Combust.~Sci.~Tech., 160, 119

\bibitem[{{Kolmogorov}(1962)}]{Kolmogorov}
{Kolmogorov}, A.~N. 1962, Journal of Fluid Mechanics, 13, 82

\bibitem[{{Lisewski} {et~al.}(2000)}]{LisHille00}
{Lisewski}, A.M., {Hillebrandt}, W., \& {Woosley}, S.~E. 2000, \apj, 538, 831

\bibitem[{{Mazzali} {et~al.}(2007)}]{Mazzali07}
{Mazzali}, P.~A., {R{\"o}pke}, F.~K., {Benetti}, S. \&
{Hillebrandt}, W. 2007, Science, 315, 825

\bibitem[{{Niemeyer}(1999)}]{Niemeyer}
{Niemeyer}, J.~C. 1999, \apjl, 523, L57

\bibitem[{{Niemeyer} \& {Kerstein}(1997)}]{Niemeyer3}
{Niemeyer}, J.~C. \& {Kerstein}, A.~R. 1997, New Astronomy, 2, 239

\bibitem[{{Niemeyer} \& {Woosley}(1997)}]{Niemeyer2}
{Niemeyer}, J.~C. \& {Woosley}, S.~E. 1997, \apj, 475, 740

\bibitem[{{Kritsuk} {et~al.}(2007){Kritsuk}, {Padoan}, {Wagner},
  \& {Norman}}]{KritPad07}
{Kritsuk}, A.~G., {Padoan}, P., {Wagner}, R., \& {Norman}, M.~L.
  2007, in American Institute of Physics Conference Series, Vol.
  932, Turbulence and Nonlinear Processes in Astrophysical Plasmas, ed.
  D.~{Shaikh} \& G.~P. {Zank}, 393

\bibitem[{{Oboukhov}(1962)}]{Oboukhov}
{Oboukhov}, A.~M. 1962, Journal of Fluid Mechanics, 13, 77

\bibitem[{{Pan} {et~al.}(2008)}]{Pan}
{Pan}, L., {Wheeler}, J.~C., \& {Scalo}, J. 2008, \apj, 681, 470

\bibitem[{{Peters}(2000)}]{Peters00}
{Peters}, N. 2000, Turbulent Combustion by Norbert Peters, pp.\ 320. Cambridge University Press

\bibitem[{{Phillips} {et~al.}(2007)}]{Phil07}
{Phillips}, M.~M., et~al.\ 2007, \pasp, 119, 360

\bibitem[{{Reinecke} {et~al.}(2002)}]{Rein02}
{Reinecke}, M., {Hillebrandt}, W., \& {Niemeyer}, J.~C.
2002, \aap, 391, 1167

\bibitem[{{R\"opke}(2007)}]{Roep07}
{R\"opke}, F.~K. 2007, \apj, 668, 1103

\bibitem[{{R{\"o}pke} \& {Hillebrandt}(2005)}]{Roep05}
{R{\"o}pke}, F.~K., \& {Hillebrandt}, W. 2005, \aap, 431,
  635

\bibitem[{{R{\"o}pke} \& {Niemeyer}(2005)}]{RoepNie07}
{R{\"o}pke}, F.~K., \& {Niemeyer}, J.~C. 2007, \aap, 464,
  683

\bibitem[{{R{\"o}pke} {et~al.}(2007){R{\"o}pke}, {Hillebrandt}, {Schmidt},
  {Niemeyer}, {Blinnikov}, \& {Mazzali}}]{Roepke}
{R{\"o}pke}, F.~K., {Hillebrandt}, W., {Schmidt}, W., {et~al.} 2007, \apj, 668,
  1132

\bibitem[{{Schmidt}(2007)}]{Schm07}
{Schmidt}, W. 2007, \aap, 465, 263

\bibitem[{{Schmidt} {et~al.}(2009){Schmidt}, {Federrath}, {Hupp}, {Kern}, \& {Niemeyer}}]{SchmFeder09}
{Schmidt}, W., {Federrath}, C., {Hupp}, M., {Kern}, S., \& {Niemeyer}, J.~C.
2009, \aap, 494, 127

\bibitem[{{Schmidt} {et~al.}(2008{\natexlab{b}}){Schmidt}, {Federrath}, \&
  {Klessen}}]{SchmFeder08}
{Schmidt}, W., {Federrath}, C., \& {Klessen}, R. 2008, Physical Review Letters, 101, 194505

\bibitem[{{Schmidt} {et~al.}(2006)}]{SchmNie06a}
{Schmidt}, W., {Niemeyer}, J.~C., \& {Hillebrandt}, W. 2006, \aap, 450, 256

\bibitem[{{Schmidt} {et~al.}(2006)}]{SchmNie06b}
{Schmidt}, W., {Niemeyer}, J.~C., {Hillebrandt}, W., \& {R\"opke}, F.~K. 2006, \aap, 450, 283

\bibitem[{{Schmidt} \& {Niemeyer}(2006)}]{SchmNie06}
{Schmidt}, W., \& {Niemeyer}, J.~C. 2006, \aap, 446, 627

\bibitem[{{She} \& {Leveque}(1994)}]{She}
{She}, Z.-S. \& {Leveque}, E. 1994, Physical Review Letters, 72, 336

\bibitem[{{She} \& {Waymire}(1995)}]{She2}
{She}, Z.-S. \& {Waymire}, E.~C. 1995, Physical Review Letters, 74, 262

\bibitem[{{Timmes} \& {Woosley}(1992)}]{Timmes}
{Timmes}, F.~X. \& {Woosley}, S.~E. 1992, \apj, 396, 649

\bibitem[{{Woosley}(2007)}]{Woosley07}
{Woosley}, S.~E. 2007, \apj, 668, 1109

\bibitem[{{Woosley}(2009)}]{WoosPriv}
{Woosley}, S.~E. 2009, private communication

\bibitem[{{Woosley} {et~al.}(2009)}]{WoosKer09}
{Woosley}, S.~E., {Kerstein}, A.~R., {Sankaran}, V. \& {R\"opke}, F. 2009, \apj, submitted (arXiv:0811.3610)

\bibitem[{{Woosley} \& {Weaver}(1994)}]{WoosWeav94}
{Woosley}, S.~E. \& {Weaver}, T. A. 1994, in Les Houches, Session LIV, Supernovae, ed.\ S.~A.~Bludman, R.~Mochkovitch, \& J.~Zinn-Justin (Amsterdam: North-Holland), 63

\end{thebibliography}
\end{document}